\documentclass[Afour,sageapa,times]{sagej}

\usepackage{moreverb,url}

\usepackage[colorlinks,bookmarksopen,bookmarksnumbered,citecolor=red,urlcolor=red]{hyperref}

\usepackage{apacite}  
\usepackage{xcolor}  
\usepackage[caption=false]{subfig} 
\usepackage{booktabs}  
\usepackage{multirow}  
\usepackage[export]{adjustbox}  

\usepackage{soul} 
\usepackage{comment} 

\usepackage[perpage]{footmisc}

\newenvironment{ethics}[1]%
{\subsection*{\normalsize\sagesf\bfseries Ethics}\begin{refsize}\noindent #1}%
{\end{refsize}}

\newcommand\BibTeX{{\rmfamily B\kern-.05em \textsc{i\kern-.025em b}\kern-.08em
T\kern-.1667em\lower.7ex\hbox{E}\kern-.125emX}}

\def\volumeyear{2016}

\newcommand{\hashtag}[1]{\texttt{\##1}}
\newcommand{\bb}[1]{\textbf{#1}}  
\newcommand{\ii}[1]{\textit{#1}}  
\newcommand{\mcc}[1]{\multicolumn{2}{c}{#1}}
\newcommand{\mccc}[1]{\multicolumn{3}{c}{#1}}

\begin{document}

\runninghead{Nasim et al.}

\title{Are we always in strife? A longitudinal study of the echo chamber effect in the Australian Twittersphere}

 \author{
     Mehwish Nasim\affilnum{1,2,4,*}, 
     Derek Weber\affilnum{2,5,*}, 
     Tobin South\affilnum{2,6}, 
     Jonathan Tuke\affilnum{2}, 
     Nigel Bean\affilnum{2}, 
     Lucia Falzon\affilnum{3} 
     and 
     Lewis Mitchell\affilnum{2,4} 
 }

 \affiliation{\\
 \affilnum{*}MN and DW contributed equally to this work.\\
     \affilnum{1}Flinders University, Australia\\
     \affilnum{2}The University of Adelaide, Australia\\
     \affilnum{3}University of Melbourne, Australia\\
     \affilnum{4}ARC Centre of Excellence for Mathematical and Statistical Frontiers, Australia\\
     \affilnum{5}Defence Science and Technology Group, Australia\\
     \affilnum{6}Massachusetts Institute of Technology
 }

 \corrauth{
     Mehwish Nasim, 
     Flinders University
     College of Science and Engineering,
     Tonsley,
     South Australia, Australia.
 }

 \email{mehwish.nasim@flinders.edu.au}

\begin{abstract}

Contrary to expectations that the increased connectivity offered by the internet and particularly Online Social Networks (OSNs) would result in broad consensus on contentious issues, we instead frequently observe the formation of polarised echo chambers, in which only one side of an argument is entertained. 

These can progress to filter bubbles, actively filtering contrasting opinions, resulting in vulnerability to misinformation and increased polarisation on social and political issues. These have real world effects when they spread offline, such as vaccine hesitation and violence. 
This work seeks to develop a better understanding of how echo chambers manifest in different discussions dealing with different issues over an extended period of time.

We explore the activities of two groups of polarised accounts across three Twitter discussions in the Australian context. 
We found Australian Twitter accounts arguing against marriage equality in 2017 were more likely to support the notion that arsonists were the primary cause of the 2019/2020 Australian bushfires, and those supporting marriage equality argued against that arson narrative. We also found strong evidence that the stance people took on marriage equality in 2017 did not predict their political stance in discussions around the Australian federal election two years later. 
Although mostly isolated from each other, we observe that in certain situations the polarised groups may interact with the broader community, which offers hope that the echo chambers may be reduced with concerted outreach to members.

\end{abstract}


\maketitle

%
%
%

\section{Introduction}
\label{sec:intro}

The increased connectivity and relative anonymity offered by the internet and especially by social media platforms (aka online social networks, or OSNs) was once hoped to provide a mechanism for a more inclusive society, especially with regard to political involvement, “promot[ing] more civic engagement and participation in elections” \cite<p.40,>{Hwang2012}. OSNs in particular allow people to connect with friends, family and like-minded individuals to form and maintain communities with shared beliefs, values and interests. Observers of modern social media will note, however, that, like with any complex system, there are unintended consequences of making reaching out to others so easy, including the broad spread of conspiracies (e.g., QAnon and the Flat Earth Society) \cite{soufan2021qanon,brazil2020flatearth}, increased polarisation \cite{Garimella2017}, especially in political discussions \cite{Garimella2018echo}, providing environments for radicalization \cite{Badawy2018} and extremism \cite{Baumann2021}, and coordinated aggression \cite{botsentinel2021meghan,MaricontiSBCKLS2019cscw}. The general consensus on contentious issues expected by classical opinion modelling theory \cite{DeGroot1974,Baronchelli2018} has instead been replaced by communities focused around competing stances on those issues, \emph{echo chambers} in which only one opinion is entertained \cite{Pariser2012,Bruns2019}, entrenched by online recommender systems preventing contrary voices from entering, thus forming \emph{filter bubbles} \cite{Pariser2012}, which leaves us vulnerable to misinformation \cite{Nikolov2021} and disinformation \cite{Starbird2019}. When this misinformed aggression moves beyond the online sphere it has real world effects such as vaccine hesitancy and anti-lockdown movements in a time of pandemics \cite{Broniatowski2018,loucaides2021,Loomba2021}, and violence \cite{samuels2020india}, some of which is politically motivated \cite{Scott2021capitolriots,mackintosh2021eth}.

The dynamics of these echo chambers is of particular interest, because their entrenchment of particular viewpoints drives the in-group/out-group mentality behind polarisation, which, left unchecked, can lead to fundamental difficulties in cooperation, with particular implications for democratic political systems \cite{Bail2018}. Not all are convinced of their danger, however \cite{Bruns2019}, because individuals are known to be members of many social circles, each with their own common attributes and interests \cite<e.g., family, friends, work, or sports, referred to as \emph{foci} by>{Feld1981}, and each of these circles will provide new and potentially contrasting viewpoints on a variety of overlapping issues. Questions remain over how these social circles and echo chambers influence social behaviour, both online and offline \cite{Bruns2019,nasim2019asnac}, but it is known that there is alignment between some sets of opinions \cite{Baumann2021}, particularly with regard to political viewpoint \cite{Jost2003,Jost2017}.

Given the relative youth of OSNs, longitudinal studies of online polarisation are only just beginning to appear, but often seek to follow polarisation on specific contentious issues over time \cite{Garimella2017,Garimella2017websci}. Our focus, instead, is on investigating communities that remain polarised over time across a variety of issues. Furthermore, it is important to study their activities in the context of the broader discussion to determine not just to what degree the groups isolate themselves from each other, but also how isolated the groups remain from the surrounding community. For these reasons, we require datasets in which known polarised groups are known to be active that are collected over a reasonable period of time \emph{and} relate to a variety of discussion topics. The issue of political alignment is also relevant, due to vulnerability to misinformation introduced by increased partisanship \cite{Nikolov2021} and the fact that political alignment has been observed to correlate with different personal values \cite{Jost2003}, for example, right-aligned people value tradition more than left-aligned people while left-aligned people value egalitarianism more \cite{Jost2017}.

Although OSNs share many features \cite{WeberNMF2021reliability}, the openness of micro-blog platforms, such as Twitter, Parler and Gab, where one account can directly connect to any other (via, e.g., mentions, replies and retweets and their equivalents), provides the best opportunity for accounts in polarised communities to bridge the gaps.
Doing so enables new and different information to flow between the communities, enabling the potential to grow consensus. 
In contrast, participants in Facebook, Instagram, Reddit and WhatsApp discussions can usually only refer to others in the same discussion thread or channel.
We use Twitter data in this study, as it is the longest established of the three microblogs mentioned, and has the largest and most representative user base. It also provides a freely available rich data model, which includes information on the directed interactions between accounts, resulting in an up-to-date window into the direction and degree of information and influence flow between Twitter accounts \cite{WeberNMF2021reliability}.

In this work, we examine the roles played online by members of two 
identified polarised communities in the context of three separate online discussions, each focused on different topics and themes, over the period of almost a year. The polarised groups had been identified in discussions of contentious issues:
\begin{itemize}
    \item Those using \hashtag{VoteYes} and those using \hashtag{VoteNo} (mutually exclusively) during the same sex marriage (SSM) debate during the Australian postal survey on the matter in late $2017$ \cite{nasim2019netsci}, dubbed the \emph{YES} and \emph{NO} communities, respectively; and 
    \item Those debating the role of arson and climate change during the $2019$/$2020$ Australian bushfires \cite{weber2020arsonemergency}, in which \emph{Supporters} of the arson theory were countered by an \emph{Opposer} community.
\end{itemize}
Notably, we have found these polarised groups to overlap and, at times, align, in the three datasets inspected. Our aim is to study the activities of these groups over time in different contexts to determine whether they remain polarised, and to characterise the nature of that polarisation using network and content analysis. Our network analysis relies upon accounts' interactions (i.e., retweets, replies, mentions and quotes) and the associations between topics they discuss as represented by partisan hashtags as proxies for clear stances on the issues at hand. 

\subsection{Research questions}

We will guide our investigation of these groups' behaviour with the following research questions:
\begin{description}
    \item[RQ1] Do polarised accounts continue to be active in the Australian Twittersphere over a period of years?
    \item[RQ2] Is their polarisation reflected in a range of their interactions (on Twitter) and discussion topics, or is it limited to just a particular type of interaction?
    \item[RQ3] Are accounts found to be polarised in one dataset still polarised in later datasets, including ones discussing different topics? In particular is there any alignment between partisan communities and those that were found to be polarised over other issues (e.g., SSM, bushfires)? 
\end{description}

Our expectation is that the Australian Twittersphere is sufficiently well established to support persistent communities of accounts over long periods of time, ones which discuss related issues, and though they may be polarised on some issues, that polarisation may not be so pronounced on others and the communities may, at times, overlap. If this is found to be true, we can conjecture that the filter bubble effect is not as strong as it was thought to be, and the echo chambers constantly reconfigure and reorganise, allowing interaction between the members of different communities. Such an observation will also be inline with previous social interaction theories that established that people are a part of various overlapping social circles \cite{Feld1981}. 

We also expect that the degree of polarisation will vary across interaction types because different interaction types are used for different purposes. Interactions between accounts may be direct, requiring that one account be aware of the other's identity (e.g., with an @mention, a reply or retweet), while others are indirect, requiring only knowledge of intermediary data and perhaps an associated common stance (e.g., common use of a partisan hashtag or URL). For direct interactions, there is the possibility that the connection is made because of a personal connection (e.g., a friendship or indication of personal respect) in addition to an agreement on stance. Furthermore, different direct interactions have different audiences: while a reply or a mention may be directed at the replied to or mentioned account, a retweet or quote tweet is aimed at the poster's followers despite the reference back to the originator of the retweeted or quoted tweet. For this reason, networks built from different interactions can be expected to exhibit different degrees of polarisation.

\subsection{Contribution}

This work provides the following contributions to the literature:

\begin{enumerate}
    \item 
    Two 
    original datasets 
    on the 
    2017 SSM debate in Australia, 
    and the 2019 
    Australian federal election;\footnote{These datasets are available at: URL redacted due to double blind review 
    } 
    
    \item A methodology for the analysis of online polarisation between two non-overlapping groups based on their behaviour and discussion content; and 
    \item A longitudinal study of two sets of such polarised communities and their degree of alignment over a series of three Twitter datasets.
\end{enumerate}


\section{Related work}

Due to the breadth of related work in this field, we structure this section as follows. We first elaborate on dangers caused by allowing polarisation to flourish online, then consider the difficulties of opinion formation in real world environments where contentious issues and opinions on them abound. We then touch on related polarisation research,
before clarifying where our work makes its contribution.

\subsection{The broken promise of social media}

As mentioned, OSNs allow people to easily form communities with shared ideas, ideals and beliefs. This notion of people connecting based on similarities is known as \emph{homophily} 
\cite{rogers1970homophily,mcpherson2001homophily}.
The open nature of the internet and social media was expected to facilitate broader engagement in society, allowing ordinary folk to communicate directly with elites \cite{woolley2018us}, leading to what Habermas referred to as \emph{deliberative democracy} \cite{Habermas1996}, where people could more easily come to a consensus on issues of interest, or gain an understanding of opposing views \cite<as discussed by>{graham2017}.
Instead, social media users have found a plethora of explicit and implicit ways to use the features of OSNs beyond their intended functions. 

These include:
\begin{itemize}
    \item the creation of echo chambers and subsequent filter bubbles \cite{Pariser2012,Bruns2019} leading to opportunities for anti-social groups to form \cite{Massanari2016}, incite and radicalise their members, and conduct organised raids on other online communities \cite{DattaA19conflictnetwork,BurgessMF2016gamergate,MaricontiSBCKLS2019cscw}---this radicalisation can move offline also, ultimately resulting in terrorist attacks \cite{wargoesviral2016,crest2017,waldek2020christchurch,Scott2021capitolriots};
    \item the use of automation, big data holdings, and organised inauthentic effort to influence both domestic and foreign politics \cite{woolley2016autopower,shorey2016,king_pan_roberts_2017,Dawson2019} as far back as 2010 \cite{metaxas2012};
    \item the realisation that misinformation can be monetised (without the requirement for malicious intent exhibited by scammers), such as when Macedonian teenagers found they could generate revenue from GoogleAds when they created highly conservative but entirely fictional news articles in the lead up the US 2016 presidential election \cite{subramanian2017}; 
    \item that public support can be manufactured with the use of employees and motivated volunteers \cite{king_pan_roberts_2017,woolley2018us,Jamieson2020} or, failing that, simply faked with automated follower accounts \cite{Aggarwal2015,Aggarwal2018,followerfactory2018}; and
    \item coordinated anonymous malicious campaigns against prominent individuals \cite<e.g.,>{botsentinel2021meghan, nasim2018real}, groups \cite{Pacheco2020www,StarbirdAW2019cscw,Graham2020} or countries \cite{graham2020virus,strick2021chinaprop,schliebs2021} 
    for ideological reasons or simply for the lulz \cite{drenton2018bikinibridge,HineOCKLSSB2017kekcucks}.
\end{itemize}
Some of these dangers have been foreseen, however, as recent revelations from whistleblowers have revealed that Facebook knew an increase in inflammatory content would be a likely consequence of its policy to weight Reactions five times more than Likes \cite{merrill2021}. In some cases the OSNs have even facilitated the spread of misinformation through favourable treatment of prominent accounts \cite{timberg2021}. 
These are all are factors contributing to the increased aggression and polarisation observed not only in the online space \cite{Garimella2017}, but also offline as a direct result of those online events. These have real world effects, such as vaccine hesitancy in a global pandemic \cite{Broniatowski2018}, coordinated anti-lockdown movements \cite{loucaides2021}, extremism facilitated by conspiratorial thinking \cite{soufan2021qanon,brazil2020flatearth} and exacerbated by bias in the media \cite<e.g.,>{Barry2020mw} and its online amplification \cite{huszar2021polamp}, radicalisation \cite{Badawy2018} and violence \cite{samuels2020india,Scott2021capitolriots,mackintosh2021eth}.

\subsection{Opinion formation in a complex opinion space}

Classical opinion modelling theory tells us that, assuming people have an opinion on any matter, increased interaction will shift the population towards consensus \cite{DeGroot1974,Baronchelli2018} as people find more reasons that they are similar \cite<the concept of homophily,>{rogers1970homophily} than different. Despite the increased opportunity for interaction provided by the internet and social media, what we observe is the contrary: an increase in polarisation on some issues \cite{Garimella2017}, which then spills from one issue to sets of issues and from the online world to the offline world. 
Although it might be reasonable to assume that accounts highly polarised on certain issues are unlikely to change their stance on those issues over time, they may be more receptive to alternative viewpoints on a variety of other topics. The online opinion space is complex as it consists of many competing diverse, often incompatible, often orthogonal, sets of opinions. For example, in the recent COVID-19 coronavirus pandemic, healthcare and economics experts' opinions on lock-downs were consistently contested and undermined by conflicting information on social media and in the news media \cite{ali2020,Tasnim2020}.
Studies have shown that online polarisation can persist even across conceptually unrelated issues \cite{Haeussler2018}, and that stances on some issues seem to align \cite{Baumann2021}, which may contribute to online friendships, consolidating the social groups and the polarisation between them. These findings highlight the need to better understand how humans respond to multiple and sometimes conflicting opinions, particularly in the online context.

\subsection{Polarisation research}

Online polarisation is a broad and well studied topic \cite{garimella2018,kligler2020interpretative}, particularly in the context of politics and from a variety of analytical perspectives and disciplines.

As social media has increasingly been used for political communications, election-related discussions have become rich sources of study of polarisation \cite{bessi2016,woolley2018us,morstatter2018alt,Garimella2018echo}.
The primary fear associated with online polarisation and the echo chambers and filter bubbles that contribute to it is that, by their very nature, they may constrain the opportunities to exchange views, let alone establish common ground. They reinforce existing opinions, risking extremism and radicalisation \cite{Bruns2019,Baumann2021,Jiang2021}.

The concern around echo chambers creating filter bubbles is not shared by all, however: Bruns argues that although a group may form an echo chamber in an online space, the individuals have many opportunities to obtain information via other communication channels, both online and offline \cite{Bruns2019}. Nevertheless, online polarisation appears to be increasing \cite{Garimella2017} and can be costly to those attempting bipartisanship \cite{Garimella2018echo}. This is in part due to the dynamics between social media, and traditional and alternative media, along with political pressures \cite{Benkler2018,Jamieson2020,vanbadham2021nyt}, leaving us vulnerable to the influence of misinformation and disinformation \cite{wardle2019,Carley2020}. 



This work seeks to provide empirical evidence from an Australian perspective, providing not just a longitudinal study of opinion polarisation over a number of distinct contentious and non-contentious topics, but also considering whether polarisation extends through different methods of online interaction. Baumann et al. considered homophily and heterophily based on political opinion from surveys \citeyear{Baumann2021}, whereas we infer opinion based on users' interactions and their use of partisan hashtags, and Garimella and Weber studied polarisation on Twitter in a longitudinal setting \citeyear{Garimella2017}, but did so by focusing on particular issues rather than the communities around them.

\section{Datasets}

We analysed the following four datasets of tweets collected between 2017 and 2020. Two of those datasets were compiled on the contentious social issues of marriage equality and federal elections. 
\begin{enumerate}
    \item \textbf{Same sex marriage (SSM)} We label users based on their preferential hashtags during the Australian 2017 SSM postal survey: 
    YES accounts were those that used only \hashtag{VoteYes}, NO accounts used only \hashtag{VoteNo}, and BOTH accounts used both hashtags.
    \item \textbf{Election} A collection of tweets of the labelled users in the SSM dataset, close to the Australian federal election in 2019.
    \item \textbf{ArsonEmergency} This dataset was collected by \citeA{weber2020arsonemergency} during the 2019-2020 Australian bushfires. They found two polarised communities in the retweet network, which we refer to here as the Arson groups. One community strongly supported the arson narrative (\emph{Supporters}), claiming arson was the cause of the bushfires, while the other community opposed that narrative with fact-check articles and official announcements (\emph{Opposers}).
    \item \textbf{AFL} A three-day collection of Australian Football League (AFL) discussions was conducted over a weekend in March, 2019.
\end{enumerate}
Further information is available in Table~\ref{tab:datasets}. 
Detailed statistics and in-depth contextual information on these datasets is provided in the supplementary material. 



\begin{table*}[ht]
    \centering
    \caption{Data Statistics}
    \label{tab:datasets}
    \resizebox{\textwidth}{!}{%
        \begin{tabular}{@{}llllrrl@{}}
            \toprule
            Dataset        & Tool  & Twitter API            & Duration                   & Tweets    & Accounts & Method of Collection  \\
            \midrule
            SSM            & GNIP  & 10\% academic API      & 1 Sep to 20 Nov 2017       &  $79,725$ & $54,855$ & Keywords: \hashtag{MarriageEquality},  \\
                           &       &                        &                            &           &          & \hashtag{SSM}, \hashtag{auspol}, \hashtag{VoteYes}, \hashtag{VoteNo} \\
            Election       & TWINT & Web UI                 & 1-21 May 2019              & $398,352$ &  $4,429$ & Timeline scraping of seed accounts \\
            ArsonEmergency & Twarc & Standard Search API    & 31 Dec 2019 to 17 Jan 2020 &  $27,546$ & $12,872$ & Keyword: ArsonEmergency \\
            AFL            & RAPID & Standard Streaming API & 22-25 Mar 2019             &  $21,799$ & $11,573$ & Keyword: afl \\
            \bottomrule
        \end{tabular}
    } 
\end{table*}


\subsection{Specific Hypotheses}

We are now in a position to guide our investigation with specific hypotheses regarding these labelled groups.

Direct and indirect interactions can be expected to exhibit polarisation differently. Content-based connections made through hashtag use are based on what the hashtag expresses rather than who else is using it. For direct interactions, where the other account is known (at least by name), that other identity may influence a user's decision to interact or not. 
For these reasons, we might expect that the polarisation evident in the Arson groups might spread across other interactions (e.g., from retweets to mentions, replies and quotes) because the accounts know each other, whereas polarisation across hashtags (as themes) might be more diffuse, because they relate to opinions and are not directly associated with individuals. People's opinions (which guide their hashtag use) may have more variety and overlap differently from the individuals they interact with regularly. Thus, we may expect polarisation in one type of interaction to persist into others, but less polarisation in content as the discussion changes to different topics.

Now knowing our labelled groups, our hypotheses moving forward are that:

\begin{enumerate}
    \item Because the SSM groups are so tightly tied to the use of \hashtag{VoteYes} and \hashtag{VoteNo} and the previously mentioned strong association between political outlook at progressive issues (such as marriage equality), we expect their interactions to be moderately homophilic and their discussion topics to be strongly homophilic, as they disagree strongly on SSM and have no evidence of other socialisation in the original SSM dataset.
    \item For the Arson groups, their retweet network strongly defines their communities based on shared opinions, so we expect strong homophily to be visible in their interactions, however it may be only moderate for mention and quote networks, which can be used to refer to non-community members without much risk of engagement or confrontation (compared with a more direct reply interaction). Furthermore, given the political and, to some degree, ideological nature of the ArsonEmergency discussion, we expect the Arson groups to also remain strongly polarised in the hashtags they use.
\end{enumerate}

\section{Methods}

We used a variety of measures to uncover polarised groups in social networks, identify their extent and characterise their connectivity and their content. We did this 
by building networks of accounts linked by interactions (retweets, mentions, replies, and quotes) and the common use of partisan hashtags, and then systematically considering a variety of measures of homophily of the polarised communities within those networks.

\subsection{Constraints of OSN data}

Despite the appeal of social media as a rich data source for sociological research, a number of questions and challenges remain. For example, restricted access to OSNs' data via their Application Programming Interfaces (APIs) limits the social networks built from such data \cite{nasim2016investigating}, the retrieved data may have inconsistencies \cite{WeberNMF2021reliability}, reproducibility of results is not always possible \cite{Assenmacher2021}, and there is a lack of robust sociological theories about social media interaction \cite<e.g.,>{schroeder2018}. 



That said, interactions on social media, limited in data model though they may be, provide the best portal we have to relevant data and therefore the best opportunity to understand the degree and nature of activity between particular actors at a particular time on a given topic of discussion.

\subsection{Social networks}


Social networks of accounts can be built with a variety of information to define the edges in the network. In traditional SNA, relations are evidence of long-standing relationships between actors, such as familial or friend relations, or organisational structures, such as supervisory or collaborative relations, but online connections differ \cite{wasserman1994social,nasim2016inferring,borgatti2009network}. Beyond follower relations, which are very easy to create but very quickly can become stale, rendering them of limited meaningful value, Twitter, for example, provides no other data on long-standing relations between accounts. Instead, direct interactions between accounts, such as retweets, mentions, replies, and quotes, can provide evidence of the currency of connectivity, the degree of interaction activity and its direction, and thus we use these interactions to study the communities in these datasets. We also use hashtags as proxies for content, and consider networks of accounts that use the same hashtags, though given the prevalence of some hashtags, we constrain the set of hashtags used. 

\subsubsection{Interaction network construction}

To build an interaction network, which we define as a weighted directed network of accounts, 
$G=(V,E)$, where $V$ is the set of nodes, and $E$ is the set of edges between them, representing when they retweet, mention, quote, or reply to each other. Nodes have a label attribute for the name of the polarised group to which they belong (if they belong to no group, it is given the value `OTHER'). The frequency of the interactions in the dataset is recorded as the weight of the edge.



\subsubsection{Hashtag co-mention networks}
The \emph{hashtag co-mention (account) network} also consists of nodes representing accounts, but they are linked with an undirected weighted edge when the accounts use the same hashtag. The edge weights are the sums of the product of the number of uses each account made of a given hashtag, for each hashtag they both used. So, for example, if accounts $\{u, v \in V\}$ use a set of common hashtags, $\{ h_1, h_2, \ldots h_n \in H \}$, we create an undirected edge $\{u,v\}$. If $h_i^u$ indicates how often user $u$ used hashtag $h_i$, the weight of the new edge is the given by 

\begin{equation}
    w_{\{u,v\}} = \displaystyle\sum\limits_{i=0}^n h_i^u \cdot h_i^v.
\end{equation}

Others \cite<e.g.,>{magelinski2021} use the minimum of $u$ and $v$'s usages of each hashtag, but their aim was to reduce computational overheads, whereas our datasets are small enough that that is not a limitation. Instead our weight calculation emphasises links from quiet (i.e., those with a small number of uses of a hashtag) accounts to loud accounts (i.e., ones with many uses), highlighting links that might otherwise be obscured or filtered out.

Some hashtags appear frequently in social media datasets, especially ones used as query terms to create the dataset in the first place (in which case it may appear in every single post). Creating a hashtag co-mention network using such popular hashtags will result in a very dense network in which many edges may lack any significant meaning. Instead, we can examine the distribution of hashtag use in a dataset and remove the most widely used hashtags. 

Further meaningful filtering can be employed by considering the content of the hashtags; in political datasets, partisan hashtags are usually indicative of (1) an opinion on an issue (2) that potentially creates an axis of polarisation depending on how strongly it divides accounts, and (3) an association with one of the polarised groups. 



\begin{figure}[t]
    \centering
    \includegraphics[width=0.99\columnwidth]{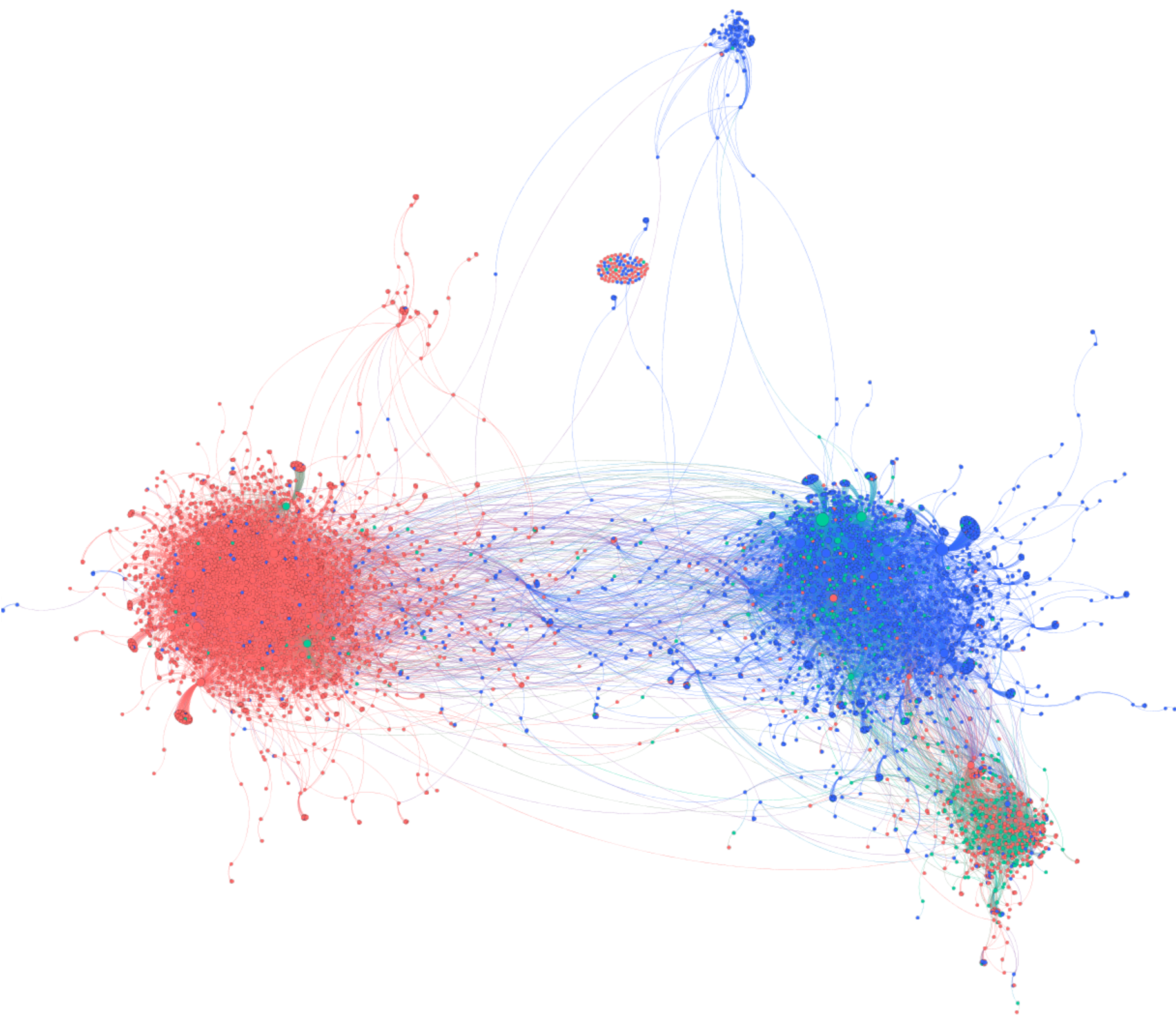}
    \caption{The largest component of the network of follow relations of the YES (blue), NO (red) and BOTH (green) accounts. The directed edges are coloured according to the following (i.e., source) node. Although BOTH accounts are primarily embedded in the YES community, the YES and NO communities are clearly polarised. (Visualised with Gephi.) }
    \label{fig:ssm_follower_network}
\end{figure}

\subsubsection{Structural analysis and visualisation}
Social theories of friendship indicate that not all ties are equal, and we have options to define the strength of ties in our networks. For networks based on interactions and content, it is possible to use frequencies as edge weights, but agnostic of the edge semantics, we can use the quadrilateral Simmelean backbone to identify the strongest ties in a given social network \cite{nick2013simmelian,nocaj2014untangling}. This approach gives high weight to edges embedded in cycles of length $4$. The intuition behind this approach is that dyads that share more common neighbours \cite<meaning they are part of a triangle, $K_3$, or cycle, $C_4$,>{Nastos2013} 
are more strongly tied -- this weight is therefore referred to as the \emph{backbone strength} of the edge. This can be used in the rendering of edges, but also the layout of network nodes.

\subsection{Homophily metrics}

Beyond simple frequency metrics of numbers of accounts, interactions, and comparisons of internal to external connection counts (i.e., how many connections are between members inside a polarised group versus connections between members of different groups), we rely on two primary measures of homophily: the assortativity coefficient \cite{Newman2003assort} and a variation on the Krackhardt E-I Index \cite{Krackhardt1988}. Assortativity is the tendency for nodes to be connected to similar nodes, for a given value of `similar', similar to homophily but agnostic of network semantics. Here, it is defined by the node's label attribute. This measure makes no use of edge weights. The Krackhardt E\nobreakdash-I Index is a simple ratio of edges internal to a community,~$I$, (i.e., between community members) and edges external to that community, $E$, (i.e., edges which have only one endpoint within the community):

\begin{equation}
    E\textnormal{-}I \; Index = \frac{|E| - |I|}{|E| + |I|}.  
\end{equation}


Our variation takes into account the weights of edges, because the weights represent the frequencies of individual interactions. This ensures that the strength of connections between nodes is considered, rather than simply their count. Both measures lie within $[-1,1]$, but their meaning is reversed: an assortativity score close to $1$ implies high polarisation, with the majority of edges connecting nodes with the same label, whereas an E-I Index of $1$ implies that all edges reach outside the group and no edge joins members of the same group. A value of $0$ for both metrics implies a balance between internal and external edges.

Binomial tests are used to test the statistical significance of the homophily measures. We consider $p$ value thresholds of $0.05$, $0.01$, $0.001$, and $0.0001$ to express the confidence in the significance.

\begin{figure}[t]
    \centering
    \includegraphics[width=0.99\columnwidth]{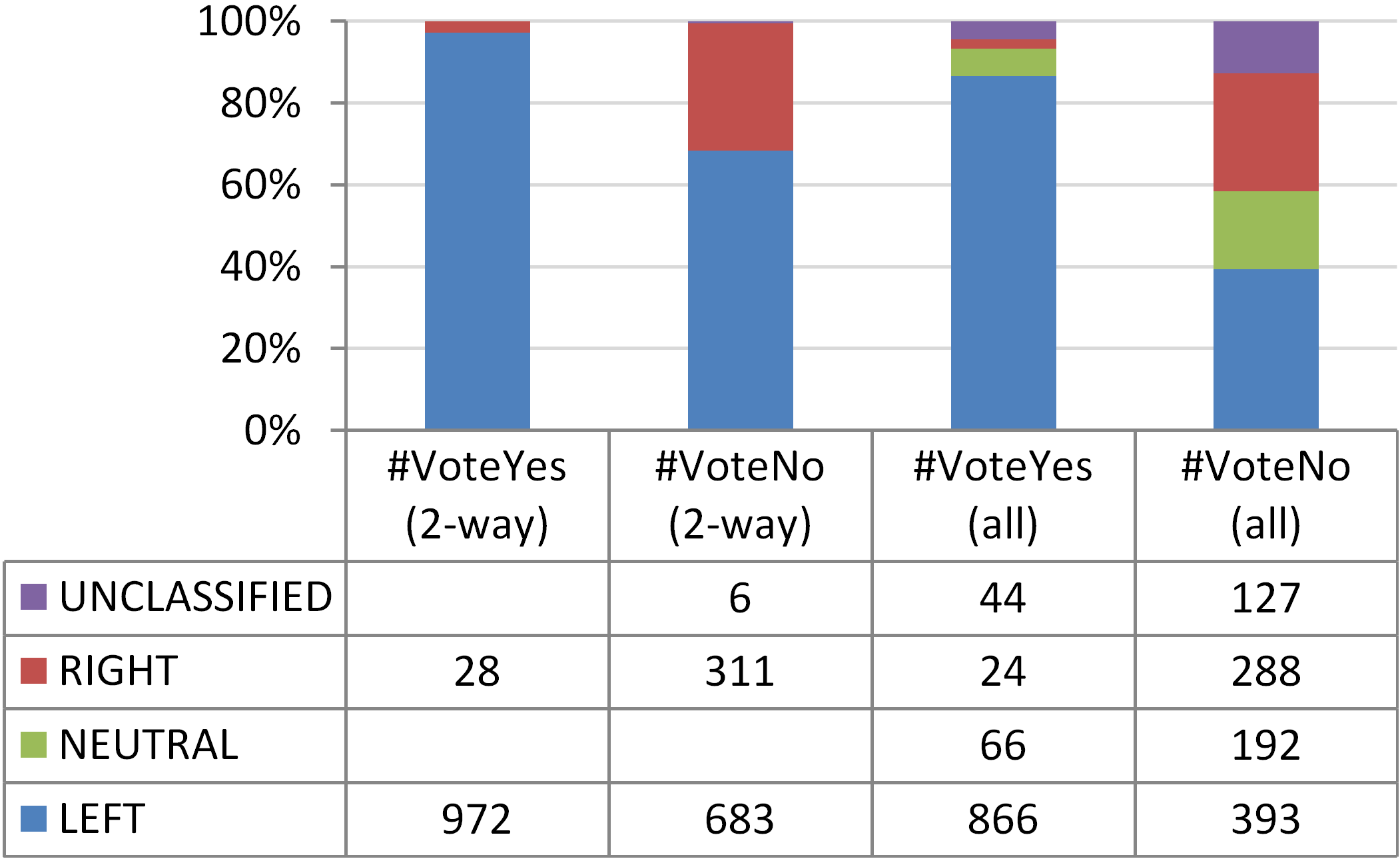}
    \caption{Results of manually labelling a random subset of tweets by YES and NO members in the Election dataset ($1{,}000$ each). The first two columns represent an attempt to assign only LEFT and RIGHT (political alignment) labels to the tweets, while in the second two columns, a further category of NEUTRAL was included. A political alignment for UNCLASSIFIED accounts could not be assigned due to a lack of suitable content. }
    \label{fig:adams_labels}
\end{figure}

\section{Results}


We address the research questions posed in the 
Introduction 
through the lens of the polarised groups identified in the SSM and Bushfires datasets. Initally, we confirm the presence of polarisation between the SSM groups \cite<polarisation in the ArsonEmergency dataset has already been confirmed by>{weber2020arsonemergency}. We then consider the activity of the polarised accounts over an extended period of ten months, whether the polarisation spans interaction types and discussion topics, and then whether the polarisation remains regardless of the topic of the discussion.

\subsection{Polarisation in the SSM discussion}
In the SSM tweets, as mentioned above, one set of hashtag filter terms were general in nature, referring to the marriage equality voting activity and politics, namely \hashtag{MarriageEquality}, \hashtag{SSM}, and \hashtag{auspol}, while the second set reflected users' opinions about marriage equality, namely \hashtag{VoteYes} and \hashtag{VoteNo}. These last two are the defining feature of YES, NO and BOTH accounts. We hypothesised a lot of repulsion between these 
YES and NO accounts, in particular; 
specifically, we anticipated relatively high structural cohesion within the groups of accounts who used the same hashtag, and relatively low cohesion among accounts who used opposite hashtags. 

%
To consider this, we retrieved as many followers of YES, NO, and BOTH accounts as possible\footnote{An account's followers may be unavailable for a variety of reasons, such as the account being protected, suspended or deleted.} and constructed a network of their follower relations, ignoring accounts outside the YES, NO and BOTH groups. The resulting network consisted of $2{,}973$ YES nodes, $3{,}417$ NO nodes, and $473$ BOTH nodes, and $22{,}139$ directed follower edges, where $\{u,v\}$ indicates that account $u$ follows account $v$. Considering only edges adjacent to a YES or NO node, we find a E-I index of $-0.84$, implying a high degree of homophily, as expected. Further confirmation of this polarisation is evident in a visualisation of the largest component of the follower network, which includes BOTH nodes (in green) for completeness, shown in Figure~\ref{fig:ssm_follower_network}. On this basis, we can confirm the YES and NO groups are polarised, as not only do they use disjoint sets of hashtags but they mostly only follow fellow community members.



Previous work have revealed alignment between people's political leaning and their support for egalitarianism and inclusivity \cite<e.g.,>{Jost2003,Jost2017,Albada2021}, so it is a reasonable to expect a similar pattern on a progressive issue, such as marriage equality. 
To examine whether the SSM groups corresponded with political alignment, a manual review of $1{,}000$ random samples from YES and NO Election tweets was conducted. Tweets were labelled at two resolutions, one aiming for a simple two-way left-wing or liberal (LEFT), or right-wing or conservative (RIGHT) alignment label, and the other also permitting NEUTRAL and UNCLASSIFIED labels. Tweets were judged on their content, and if that were not sufficiently clear, the profile of the tweet's author would be inspected (such content was preserved in the metadata of collected tweets). The results presented in Figure~\ref{fig:adams_labels} indicate that YES members were almost exclusively LEFT-aligned, while the alignment of NO members varied much more. 
On deeper inspection, many tweets could be labelled only as NEUTRAL, which is not unexpected in a wide-ranging political discussion, as they often include simple statements of fact. Furthermore, a significant number could not be reasonably classified due to a lack of content. The implications are that the NO members are much more politically diverse than the YES members, who are mostly LEFT-aligned and that the polarisation observed in their use of \hashtag{VoteYes} and \hashtag{VoteNo} may not be sustained in other political discussions.


Based on this identification of polarised YES and NO accounts, and the analysis of their political stance, we then observed those accounts' behaviour in the lead up to the 2019 Australian federal election, with the aim of testing whether their polarisation on SSM also led to polarisation over the political issues being discussed. Prior to presenting those results, however, we discuss a significant overlap between the SSM groups and the Arson groups.

\subsection{A chance finding}
\begin{figure}
    \centering
    \subfloat[Arson groups.]{
        \includegraphics[width=0.47\columnwidth,valign=c]{images/arson-retweets-newcolors.pdf}
        \vphantom{\includegraphics[width=0.47\columnwidth,valign=c]{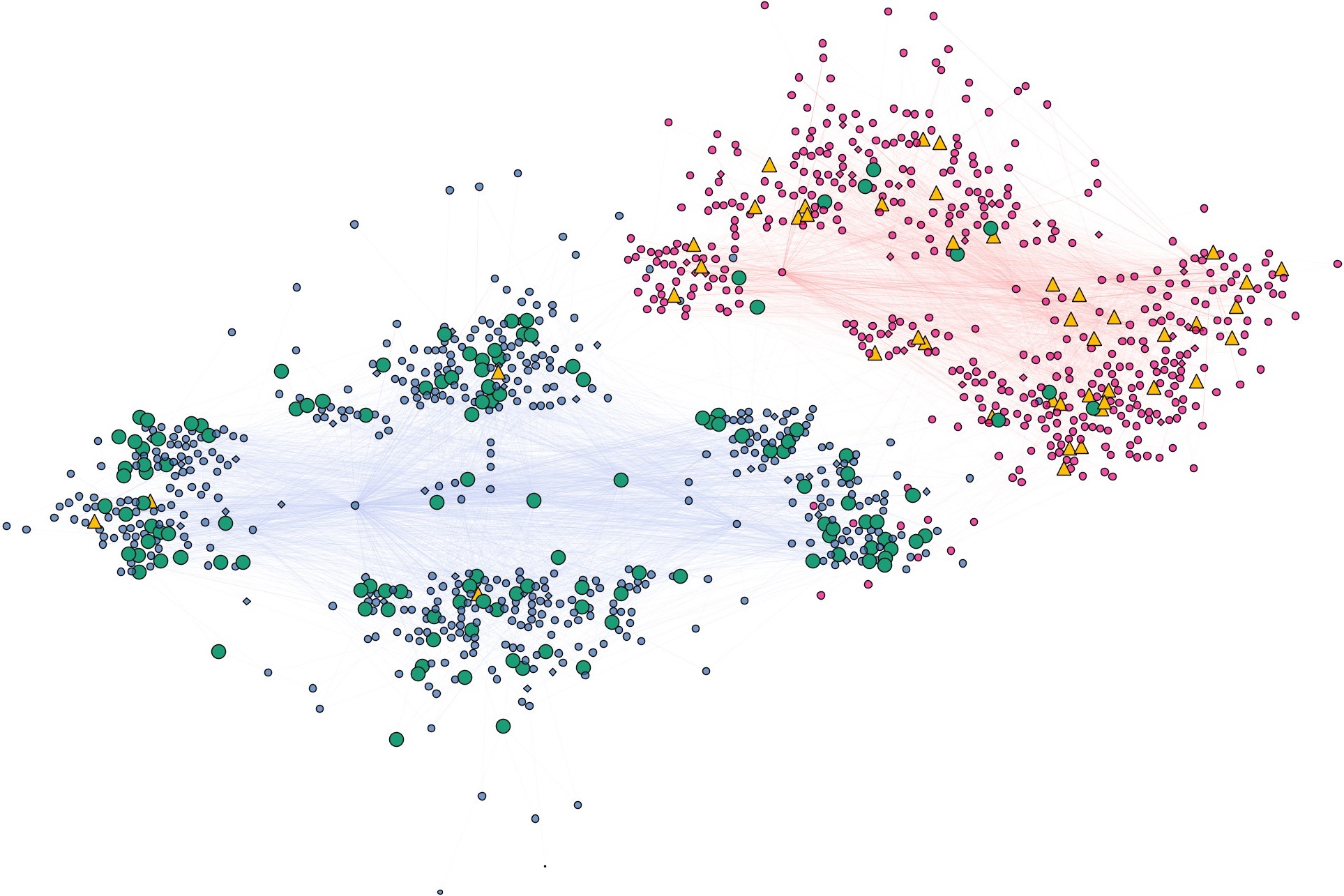}}
        \label{fig:arson_in_arson_rts}
    } 
    \subfloat[SSM groups.]{
        \includegraphics[width=0.47\columnwidth,valign=c]{images/arsonplusssm-cropped.jpg}
        \label{fig:ssm_in_arson_rts}
    }


    \caption{The retweet network in the ArsonEmergency dataset including just the Arson groups on the left \protect\cite<Left figure is reproduced with permission from>{weber2020arsonemergency}, where blue nodes are opposers of arson theory and red nodes are supporters of arson theory. The SSM groups are mapped on this figure on the right and we observed that the SSM groups appeared to have remained polarised on the arson issue; YES mapped to opposers and NO mapped to supporters. Directed edges are coloured according to their source nodes, and are semi-transparent to manage occlusion.}
    \label{fig:groups_in_arson_rts}
\end{figure}

It was observed that $1{,}015$ SSM accounts from YES, NO and BOTH groups were active in the ArsonEmergency discussion, and that they appeared to still be polarised. Furthermore, of those $1{,}015$ accounts, a full $995$ of them appeared in the retweet network, in which the Supporters and Opposers appeared. We highlighted the SSM accounts in a reproduction of the original retweet network visualisation (Figure~\ref{fig:groups_in_arson_rts}) and observed that the groups appeared to have remained polarised. To examine this statistically, one-tail probability tests for each group were used to confirm that Supporters~$\mapsto$~NO accounts and Opposers~$\mapsto$~YES accounts by rejecting the null hypothesis that the polarised groups were independent, at $\alpha = 0.01$.

With this encouragement, we used the ArsonEmergency dataset and earlier Australia-focused datasets, to determine if the SSM and Arson accounts were present and active, and whether the polarisation observed elsewhere was maintained across interaction types and in the content they discussed.

\subsection{Enduring polarisation}
There is some evidence to suggest that if people have strong moral convictions, then they are likely to continue engaging politically \cite{Skitka2008}, and so it is possible, if not likely, that those who participated in the SSM discussion and the ArsonEmergency discussion (both topics with a strong political element) would also have been active in the Australian Twittersphere in the intervening period.

We therefore now turn to examine the presence and polarisation across all four datasets. To do this, from these datasets we construct and examine retweet, reply, mention and quote interaction networks, as well as content-related networks based on hashtag co-mentions. 

\subsubsection{Continued presence}
Table~\ref{tab:everpresent_accounts} shows the number and proportion of SSM and Arson accounts in the four datasets. Although some of the proportions drop considerably from the original groups, there are still sufficient absolute numbers to draw conclusions regarding their behaviour (the smallest presence still has $42$ members, nearly $10\%$ of the original community). The considerable drop in SSM accounts, especially in the NO group, does raise the question of how these accounts have been used, despite the time between the SSM collection (late $2017$) and the earliest of the other datasets (the AFL, in early $2019$). Given so many accounts did not participate in these discussions, was it because they were still active but discussing other topics, or is it that they were used only or mostly for the SSM discussion and then left inactive. The great number of YES accounts active in the Election dataset indicates that perhaps NO accounts were used in this single-purpose manner.

\begin{table}[ht]
    \centering
    \caption{Sizes and proportions of the presence of the polarised groups in the datasets. Bolded figures belong to the original datasets.}
    \label{tab:everpresent_accounts}
    \resizebox{\columnwidth}{!}{%

        \begin{tabular}{@{}lrrrrrrrr@{}}
            \toprule
                      & \mcc{SSM}                   & \mcc{AFL}        & \mcc{Election}      & \mcc{ArsonEmergency} \\ 
                      & \mcc{\textit{Late 2017}} & \mcc{\textit{March 2019}} & \mcc{\textit{May 2019}} & \mcc{\textit{January 2020}} \\ 
                      \cmidrule(r){2-3}             \cmidrule(lr){4-5} \cmidrule(lr){6-7}    \cmidrule(l){8-9}
            YES       & \bb{8,623} & \bb{(100.0\%)} & 376 &  (4.4\%)   & 3,390    & (39.3\%) &      698 &      (8.1\%) \\
            NO        & \bb{7,880} & \bb{(100.0\%)} &  53 &  (0.7\%)   &   631    &  (8.0\%) &      148 &      (1.8\%) \\
            Supporter &         93 &       (18.7\%) &  42 &  (8.5\%)   &    72    & (14.5\%) & \bb{497} & \bb{(100\%)} \\
            Opposer   &        240 &       (40.5\%) &  73 & (12.3\%)   &   156    & (26.3\%) & \bb{593} & \bb{(100\%)} \\ 
            \bottomrule
        \end{tabular}%
        
    } 
\end{table}

\begin{table*}[ht]
    \centering
    \caption{Summary details of the inter- and intra-group interactions by the SSM and Arson polarised groups in networks built from three datasets. Significance $p$-values are based on using binomial tests with the null hypothesis that the groups had no connection preference.}
    \label{tab:summary_int_sig}
    \resizebox{\textwidth}{!}{%
        \begin{tabular}{@{}lll|rr|c||rr|c||rr|c@{}}
            \toprule
             &                           &             & \mccc{\bb{Election}}                  & \mccc{\bb{ArsonEmergency}}          & \mccc{\bb{AFL}}                     \\
             &                           &             & \mcc{\ii{Target}}       &             & \mcc{\ii{Target}}     &             & \mcc{\ii{Target}}     &             \\
             & Network                   & \ii{Source} & YES        & NO         & Sig. ($p<$) & YES        & NO       & Sig. ($p<$) & YES        & NO       & Sig. ($p<$) \\ \cmidrule{2-12}
            \multirow{8}{*}{SSM} 
             & \multirow{2}{*}{Retweets} & YES         & 55,792     & 2,359      & \ii{0.0001} & 349        & 27       & \ii{0.0001} & 45         & --       & \ii{0.0001} \\
             &                           & NO          & 2,091      & 4,677      & \ii{0.0001} & 17         & 96       & \ii{0.0001} & --         & 6        & \ii{0.05}   \\ \cmidrule{4-12}
             & \multirow{2}{*}{Mentions} & YES         & 5,337      & 209        & \ii{0.0001} & 9          & 2        & --          & 23         & 1        & \ii{0.0001} \\
             &                           & NO          & 749        & 261        & \ii{0.0001} & 24         & 21       & --          & --         & 1        & --          \\ \cmidrule{4-12}
             & \multirow{2}{*}{Replies}  & YES         & 2,231      & 106        & \ii{0.0001} & 5          & --       & --          & 21         & 1        & \ii{0.0001} \\
             &                           & NO          & 133        & 175        & \ii{0.05}   & 10         & 10       & --          & --         & --       & --          \\ \cmidrule{4-12}
             & \multirow{2}{*}{Quotes}   & YES         & 3,303      & 183        & \ii{0.0001} & 10         & 3        & --          & 10         & --       & \ii{0.01}   \\
             &                           & NO          & 250        & 335        & \ii{0.001}  & 1          & 9        & \ii{0.05}   & 1          & --       & --          \\ \cmidrule{2-12}
             & \multirow{2}{*}{Hashtags} & YES         & 42,683,122 & 79,068,551 & \ii{0.0001} & 381        & 1,258    & \ii{0.0001} & 652        & 1,577    & \ii{0.0001} \\
             &                           & NO          & 79,068,551 & 258,579    & \ii{0.0001} & 1,258      & 611      & \ii{0.0001} & 1,577      & 105      & \ii{0.0001} \\ \midrule
             &                           &             & \mcc{\ii{Target}}       &             & \mcc{\ii{Target}}     &             & \mcc{\ii{Target}}     &             \\
             & Network                   & \ii{Source} & Supporters & Opposers   & Sig. ($p<$) & Supporters & Opposers & Sig. ($p<$) & Supporters & Opposers & Sig. ($p<$) \\ \cmidrule{2-12}
            \multirow{8}{*}{Arson}
             & \multirow{2}{*}{Retweets} & Supporters  & 5,725      & 23         & \ii{0.0001} & 3,603      & 23       & \ii{0.0001} & 4          & --       & --          \\
             &                           & Opposers    & 12         & 11,878     & \ii{0.0001} & 13         & 3,006    & \ii{0.0001} & --         & 6        & \ii{0.05}   \\ \cmidrule{4-12}
             & \multirow{2}{*}{Mentions} & Supporters  & 509        & 14         & \ii{0.0001} & 567        & 343      & \ii{0.0001} & 4          & --       & --          \\
             &                           & Opposers    & 12         & 562        & \ii{0.0001} & 28         & 30       & --          & --         & 1        & --          \\ \cmidrule{4-12}
             & \multirow{2}{*}{Replies}  & Supporters  & 81         & 5          & \ii{0.0001} & 288        & 144      & \ii{0.0001} & 6          & --       & \ii{0.05}   \\
             &                           & Opposers    & 6          & 250        & \ii{0.0001} & 11         & 33       & \ii{0.01}   & --         & 1        & --          \\ \cmidrule{4-12}
             & \multirow{2}{*}{Quotes}   & Supporters  & 183        & 27         & \ii{0.0001} & 212        & 50       & \ii{0.0001} & --         & --       & --          \\
             &                           & Opposers    & 9          & 568        & \ii{0.0001} & 6          & 62       & \ii{0.0001} & --         & --       & --          \\ \cmidrule{2-12}
             & \multirow{2}{*}{Hashtags} & Supporters  & 173,488    & 977,889    & \ii{0.0001} & 106,433    & 106,922  & --          & 60         & 553      & \ii{0.0001} \\
             &                           & Opposers    & 977,889    & 5,941,413  & \ii{0.0001} & 106,922    & 7,875    & \ii{0.0001} & 553        & 37       & \ii{0.0001} \\
            \bottomrule 
        \end{tabular}%
    }
\end{table*}

\begin{figure*}[t]
	\centering
	\subfloat[Arson retweets]{
	    \includegraphics[height=0.11\textheight]{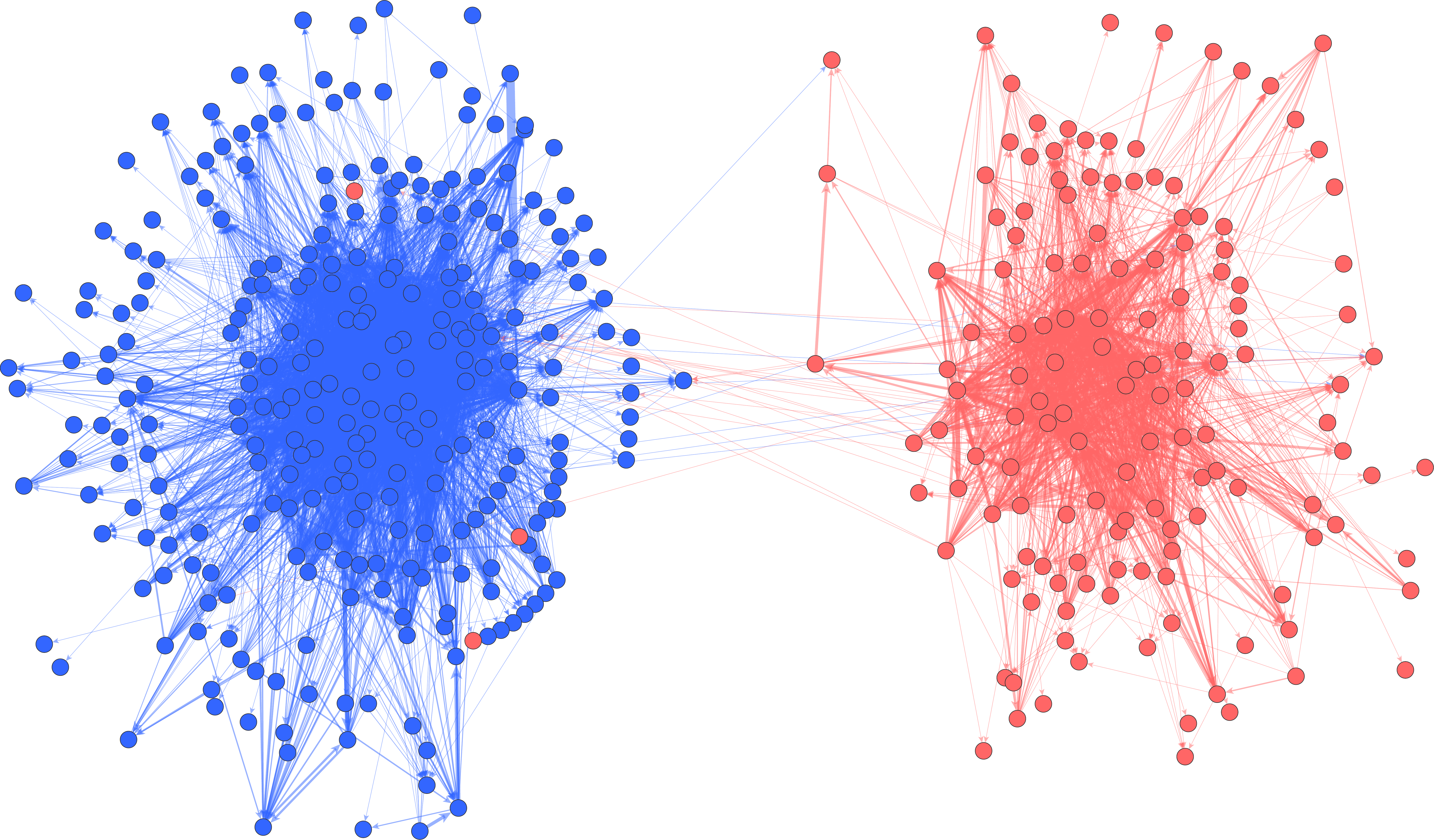}
	    \label{fig:arson_groups_in_auselecvote_retweet_network} 
	} \hfill 
	\subfloat[Arson mentions]{
	    \includegraphics[height=0.11\textheight]{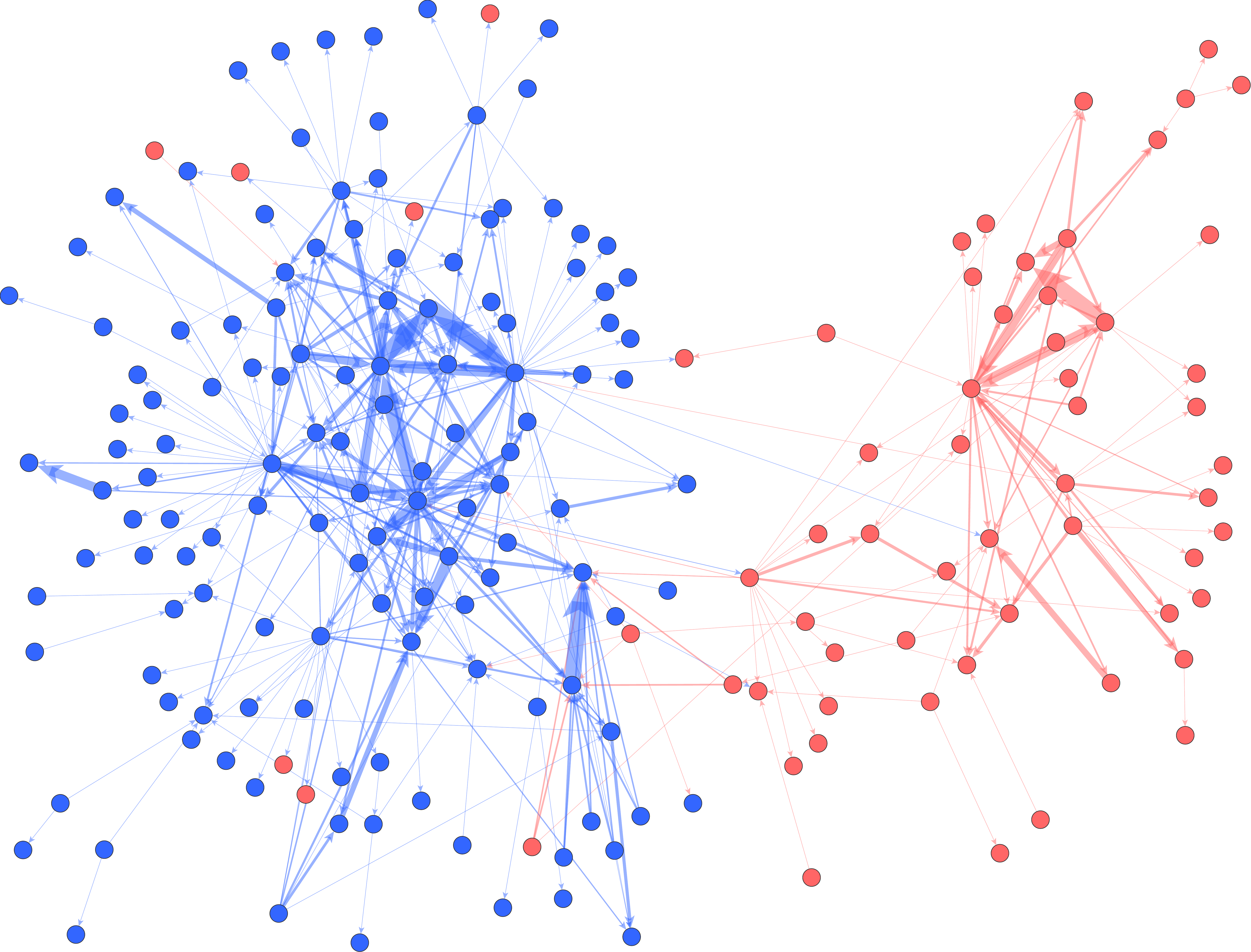} 	
	    \label{fig:arson_groups_in_auselecvote_mention_network} 
	} \hfill 
	\subfloat[Arson replies]{
	    \includegraphics[height=0.11\textheight]{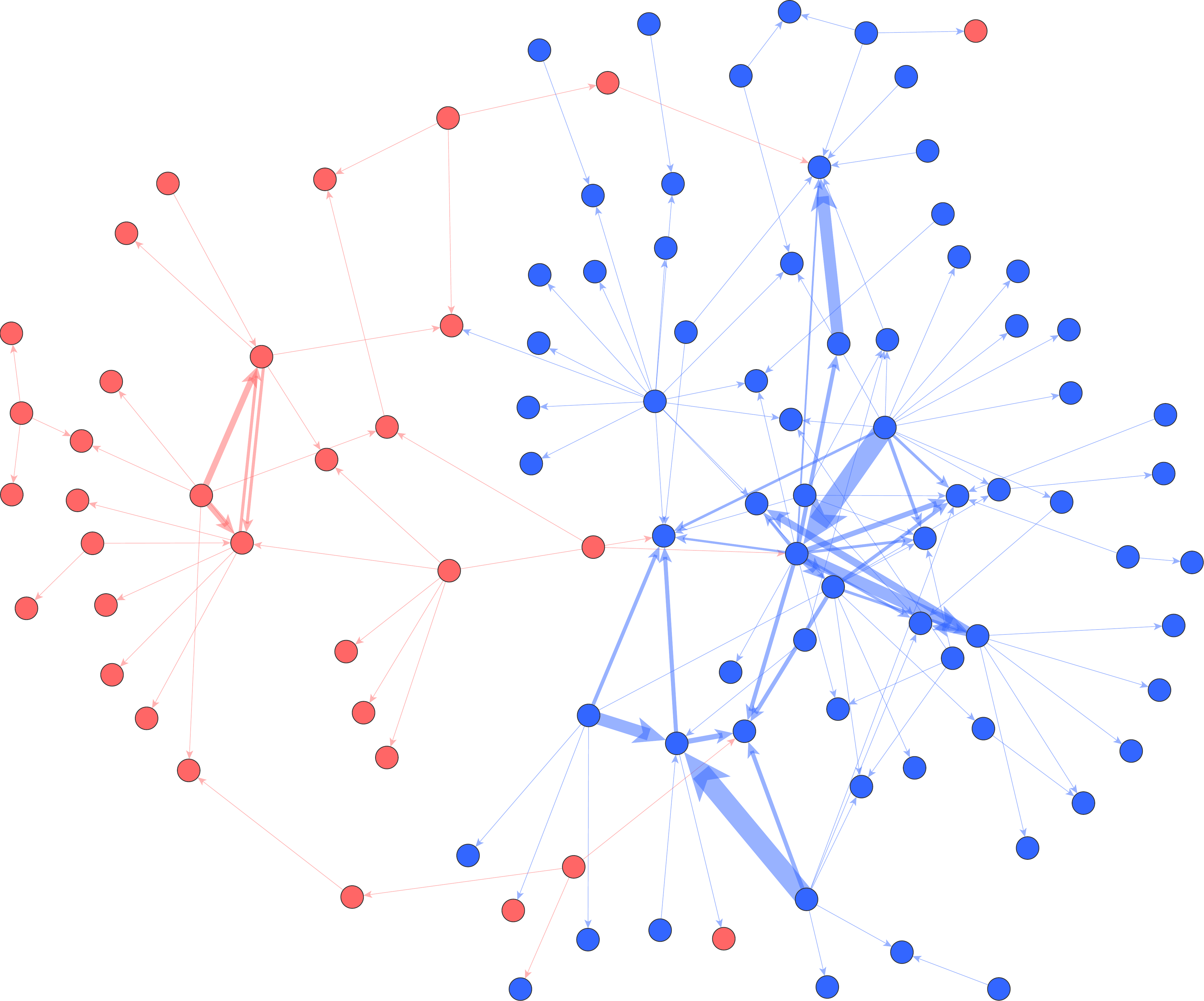} 	
	    \label{fig:arson_groups_in_auselecvote_reply_network} 
	} \hfill 
	\subfloat[Arson quotes]{
	    \includegraphics[height=0.11\textheight]{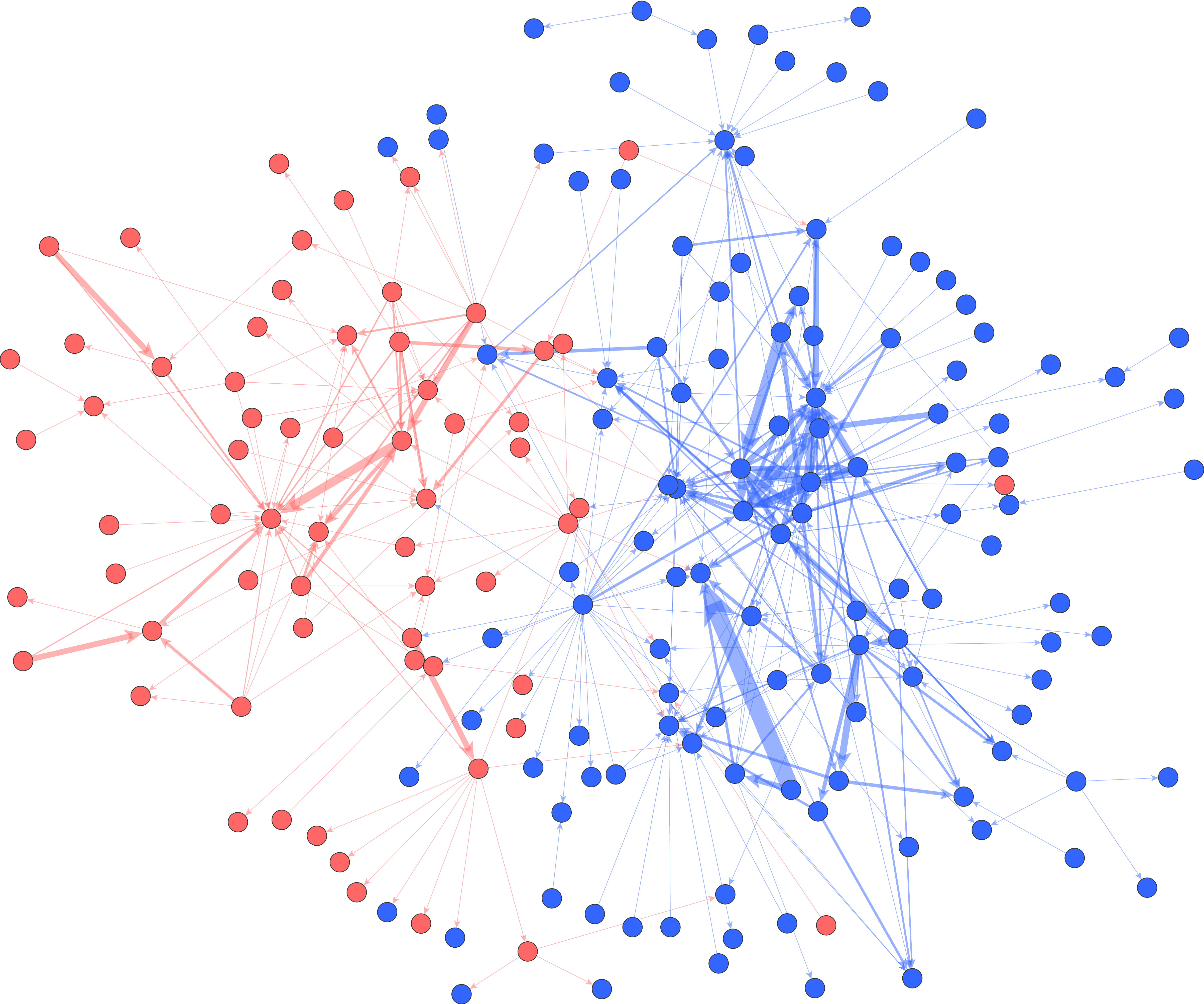} 	
	    \label{fig:arson_groups_in_auselecvote_quote_network} 
	} \\
	\subfloat[SSM retweets]{
	    \includegraphics[height=0.15\textheight]{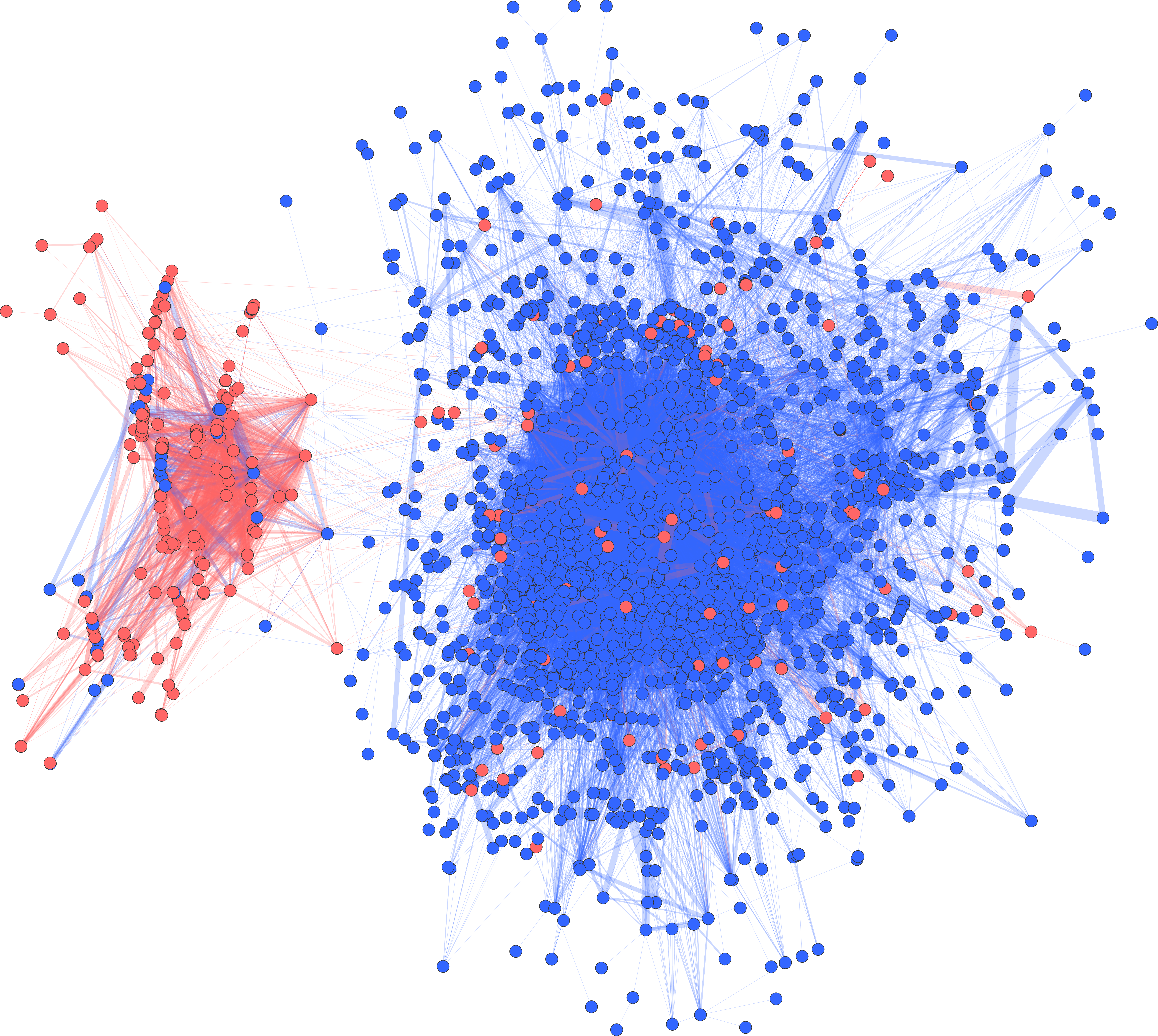}
	    \label{fig:ssm_groups_in_auselecvote_retweet_network} 
	} \hfill 
	\subfloat[SSM mentions]{
	    \includegraphics[height=0.15\textheight]{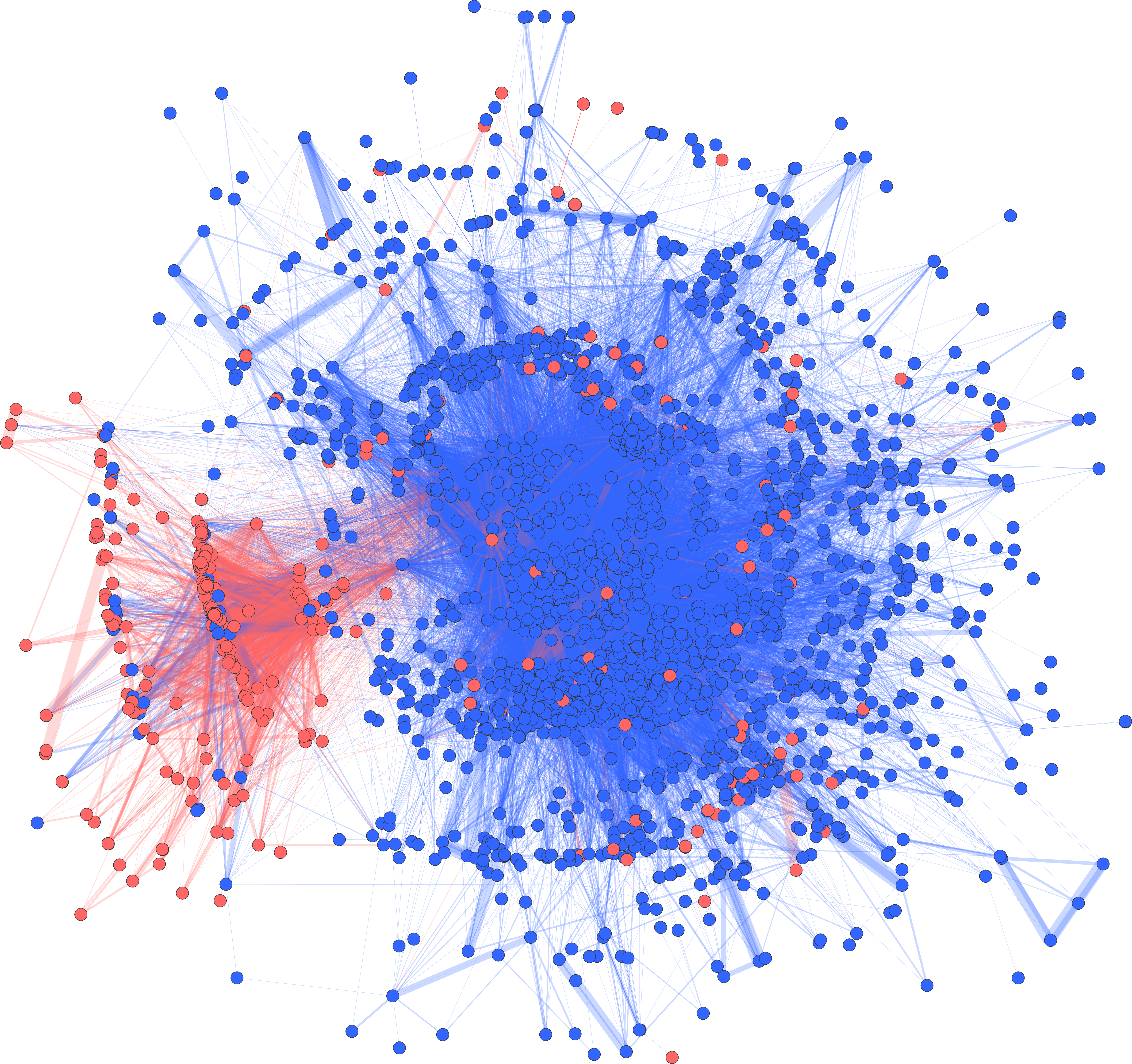} 	
	    \label{fig:ssm_groups_in_auselecvote_mention_network} 
	} \hfill 
	\subfloat[SSM replies]{
	    \includegraphics[height=0.15\textheight]{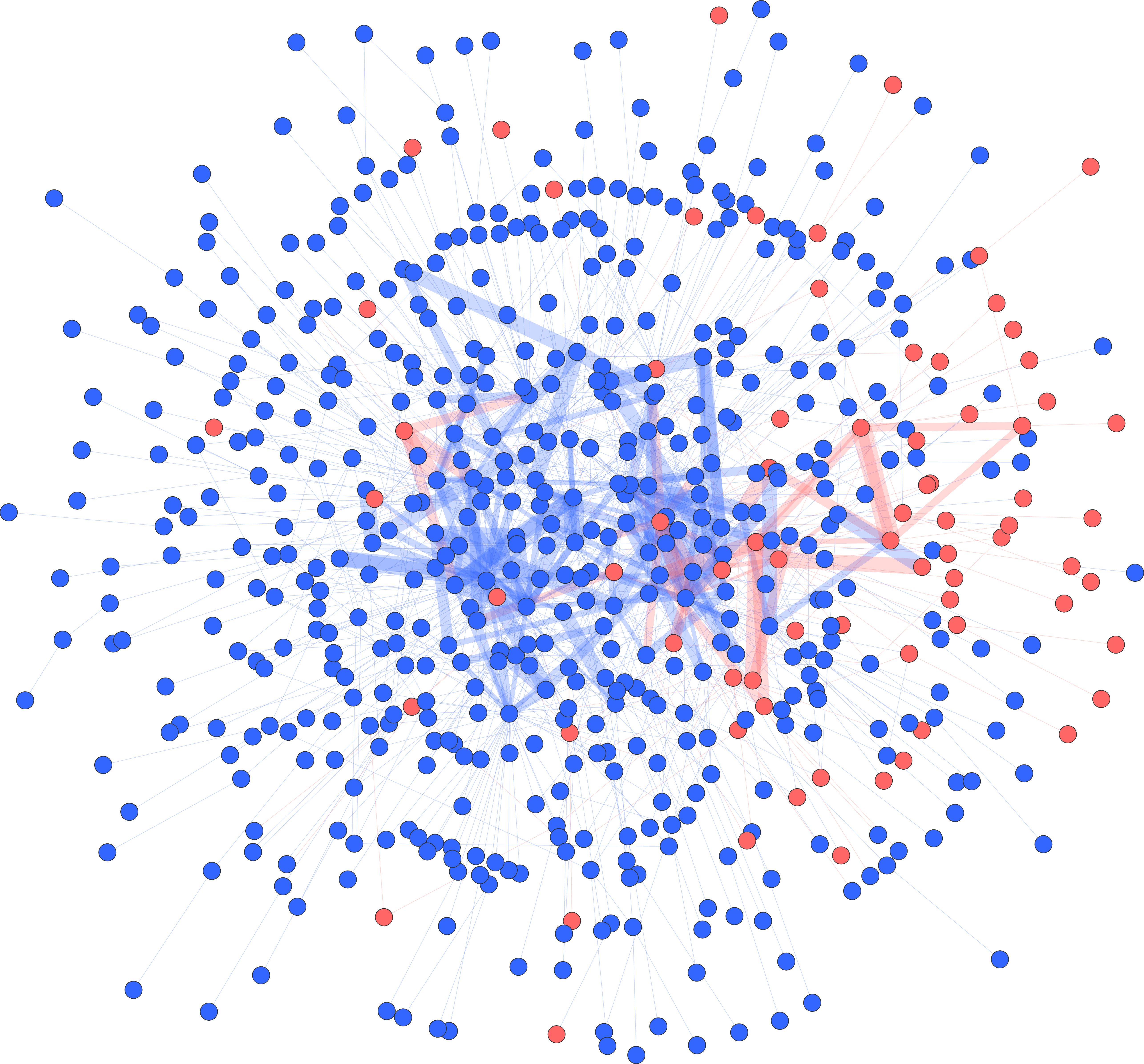} 	
	    \label{fig:ssm_groups_in_auselecvote_reply_network} 
	} \hfill 
	\subfloat[SSM quotes]{
	    \includegraphics[height=0.15\textheight]{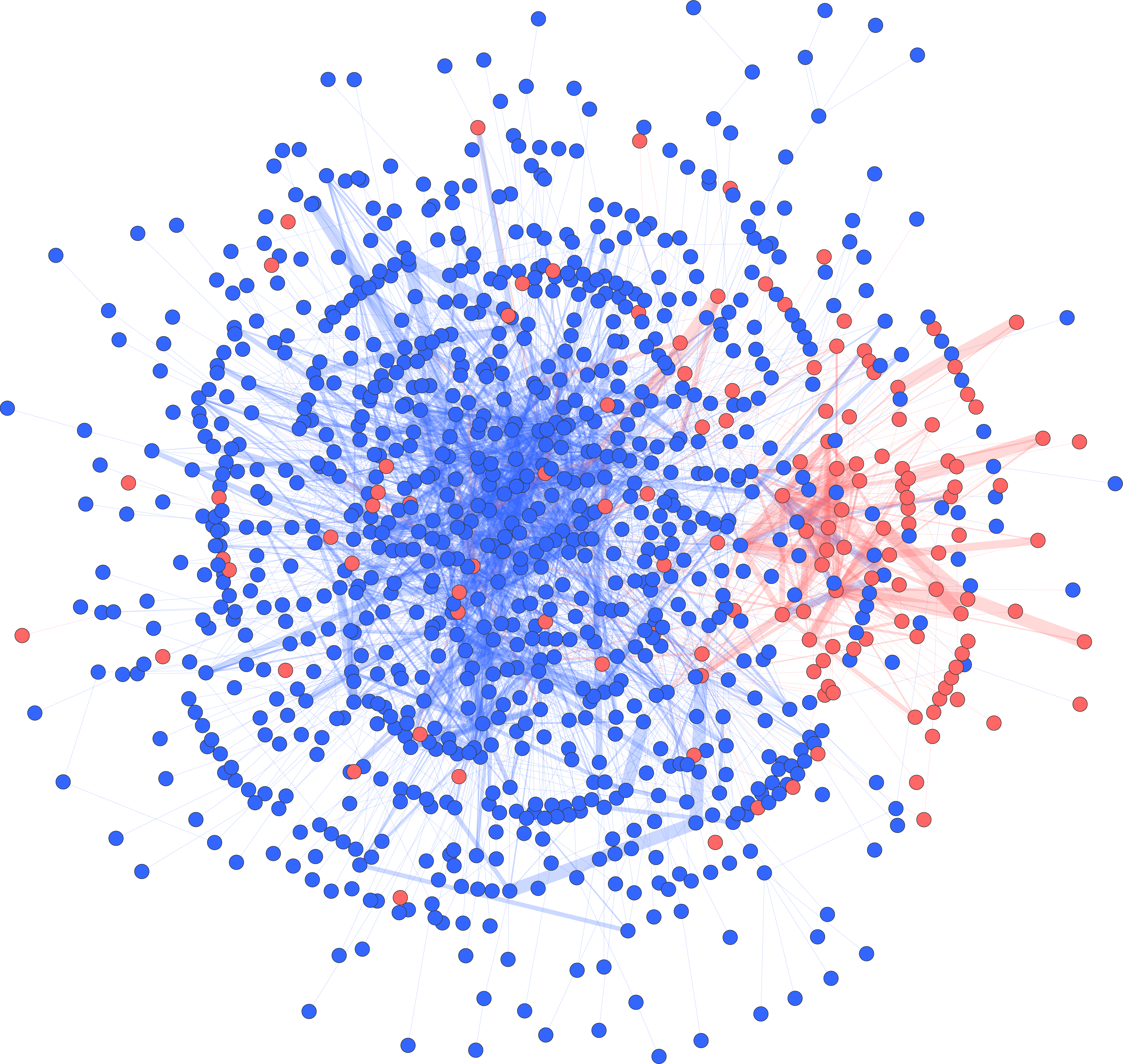} 	
	    \label{fig:ssm_groups_in_auselecvote_quote_network} 
	}
	\caption{\textbf{Polarisation in interaction networks: }The largest components in interaction networks of Supporter (red) and Opposer (blue) accounts (top row) and YES (blue) and NO (red) accounts (bottom row) active in the Election dataset. Edge width describes the backbone strength.  
	Polarised accounts from Arson and SSM datasets remain significantly polarised in elections dataset for various types of interactions.. 
	}
	\label{fig:arson_groups_in_auselecvote_networks}
\end{figure*}

\paragraph{Summary of interaction network findings}
In almost all circumstances, the echo chamber effect appears to be maintained to some degree, with internal connections preferred over external ones, especially between Supporters and Opposers. The only circumstance where that effect is reduced is in the NO groups' use of replies and mentions and Opposers' use of mentions in the ArsonEmergency dataset, where they more even in their connections. It is possible that some of these mentions were used for aggressive, rather than collegiate, interactions, but analysis of their content is required for this judgement and there were relatively few of these interactions, so any such judgement is unlikely to be indicative of a broader pattern of behaviour.

The results of a systematic examination of the presence and interactivity between the YES and NO and Supporter and Opposer accounts in the AFL, Election and ArsonEmergency datasets are presented in Table~\ref{tab:summary_int_sig}.

The polarisation in the Election dataset is statistically significant across all groups and interactions, and it is homophilic in all but one condition: NO accounts mention YES accounts more frequently than other NO accounts. This marked polarisation is also immediately apparent in visualisations of the interaction networks (Figure~\ref{fig:arson_groups_in_auselecvote_networks}).

\subsubsection{Summary of networks based on content}
Results indicate the echo chamber effect is strongly maintained across most interactions in most datasets, especially where there is reasonable amount of activity. Here we consider whether the topics also under discussion also exhibit similar patterns of polarisation, and we use hashtags as an indicator of those topics.

\begin{figure*}[ht]
    \centering
    \includegraphics[width=0.99\textwidth]{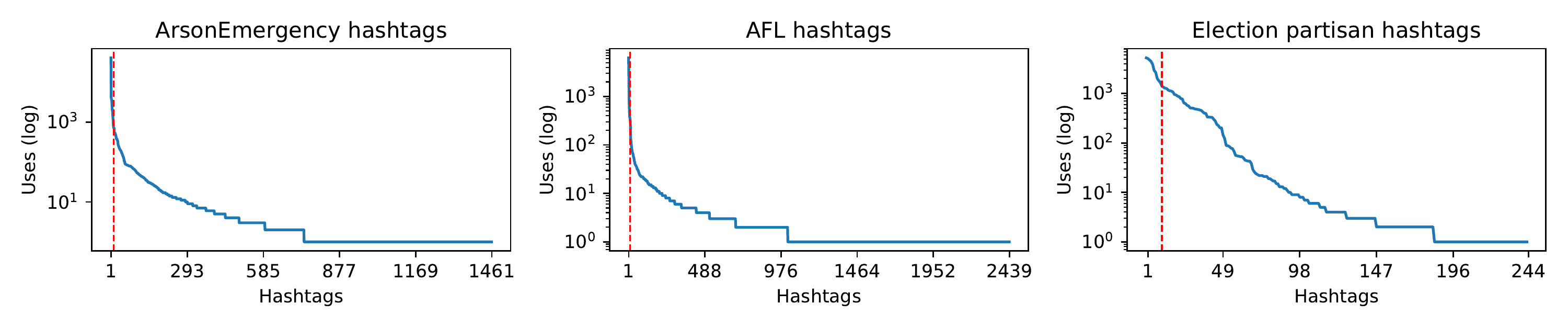}
    \caption{Hashtag use distributions for the ArsonEmergency and AFL datasets, and the distribution of the use of partisan and co-occurring hashtags for the Election dataset. The red vertical line indicates the 10th most frequently used hashtag; this and the more frequently used hashtags were removed from calculating the co-hashtag mention networks. 
    }
    \label{fig:hashtag_distributions_and_cutoffs}
\end{figure*}

First, however, we must cull the hashtags under consideration, as the high frequency of popular hashtags can hamper the discovery of the structures underlying their use. Instead, as discussed above, we explicitly filter the most frequent hashtags and we additionally make use of partisan and faux partisan hashtags. Examining the distributions of hashtag use in each of the dataset revealed that removing the ten most frequent hashtags in each would be sufficient to avoid the majority of their binding effects (shown as the dashed red vertical lines in Figure~\ref{fig:hashtag_distributions_and_cutoffs}).

Second, we developed the (faux) partisan hashtag sets. In the Election dataset, we identified $44$ hashtags of the $200$ most frequent as clearly partisan (e.g., \hashtag{corrupt<party>} or \hashtag{<party>liars}). For the AFL and ArsonEmergency datasets, we identified the ten most frequently used hashtags unique to each group. We considered the tweets containing these hashtags and created hashtag co-mention networks using all the hashtags that appeared in them (save for the most frequently occurring hashtags, as mentioned above). The number of hashtags considered for each group and dataset is shown in parentheses next to the ``Hashtags'' label in Table~\ref{tab:homophily_summary}, which also shows the number of SSM and Arson group accounts present in the resulting networks, and their respective connectivity.

\begin{table*}[ht]
    \centering
    \caption{Summary details of SSM and Arson polarised groups in networks built from three datasets. The `Hashtag' networks are the hashtag co-mention accounts, and the number in parentheses is the total count of the partisan and co-occurring hashtags.}
    \label{tab:homophily_summary}
    \resizebox{\textwidth}{!}{%
        \begin{tabular}{@{}lll|rr|rrr|rr||rrr@{}}

            \toprule
            \bb{Group}       & \bb{Dataset}   & \bb{Network}   & \mcc{\bb{Category 1/2 Nodes}} & \mccc{\bb{Category 1/2 Edge Weights}}  & \mcc{\bb{Homophily}}      & \mccc{\bb{Broader Network}}       \\ 
                             &                &                & Category 1 & Category 2       & Homophilic & Heterophilic & Sum        & E-I index & Assortativity & Nodes  & Edge Weights & E-I Index \\ \midrule
                             & Election       & Retweet        & 3,045      & 592              & 82.5\%     & 17.5\%       & 64,919     & -0.650    & 0.459         & 13,585 & 331,682      & 0.530     \\
                             &                & Mention        & 1,611      & 217              & 61.0\%     & 39.0\%       & 6,556      & -0.221    & 0.247         & 8,490  & 51,673       & 0.701     \\
                             &                & Reply          & 1,005      & 142              & 76.1\%     & 23.9\%       & 2,645      & -0.523    & 0.331         & 4,316  & 12,397       & 0.484     \\
                             &                & Quote          & 1,757      & 234              & 76.0\%     & 24.0\%       & 4,071      & -0.520    & 0.459         & 6,315  & 26,025       & 0.641     \\ \cmidrule{3-13}
            SSM              &                & Hashtags (244) & 3,390      & 631              & 53.3\%     & 46.7\%       & 47,038,810 & -0.066    & 0.029         & 4,429  & 79,327,130   & -0.143    \\ \cmidrule{2-13}
            \ii{Category 1:} & ArsonEmergency & Retweet        & 688        & 144              & 88.9\%     & 11.1\%       & 489        & -0.778    & 0.765         & 12,076 & 21,526       & 0.618     \\
            \ii{YES}         &                & Mention        & 140        & 67               & 64.2\%     & 35.8\%       & 56         & -0.285    & 0.160         & 3,206  & 7,523        & 0.881     \\
            \ii{Category 2:} &                & Reply          & 99         & 50               & 75.0\%     & 25.0\%       & 25         & -0.500    & 0.323         & 2,041  & 3,031        & 0.853     \\
            \ii{NO}          &                & Quote          & 57         & 35               & 83.5\%     & 16.5\%       & 23         & -0.669    & 0.588         & 1,268  & 1,542        & 0.779     \\ \cmidrule{3-13}
                             &                & Hashtags (171) & 698        & 148              & 98.7\%     & 1.3\%        & 998        & -0.975    & 0.970         & 12,867 & 63,819       & 0.870     \\ \cmidrule{2-13}
                             & AFL            & Retweet        & 276        & 41               & 100.0\%    & 0.0\%        & 51         & -1.000    & 1.000         & 5,735  & 7,047        & 0.761     \\
                             &                & Mention        & 177        & 19               & 97.9\%     & 2.1\%        & 25         & -0.958    & 0.648         & 7,740  & 19,222       & 0.913     \\
                             &                & Reply          & 118        & 9                & 95.5\%     & 4.5\%        & 22         & -0.909    & 0.000         & 4,815  & 6,060        & 0.745     \\
                             &                & Quote          & 78         & 7                & 50.0\%     & 50.0\%       & 11         & 0.000     & 0.000         & 1,821  & 1,670        & 0.765     \\ \cmidrule{3-13}
                             &                & Hashtags (347) & 376        & 53               & 100.0\%    & 0.0\%        & 757        & -1.000    & 1.000         & 11,573 & 76,700       & 0.871     \\ \midrule
                             & Election       & Retweet        & 158        & 314              & 99.7\%     & 0.3\%        & 17,638     & -0.995    & 0.983         & 13,585 & 331,682      & 0.643     \\
                             &                & Mention        & 120        & 243              & 97.6\%     & 2.4\%        & 1,097      & -0.952    & 0.857         & 8,490  & 51,673       & 0.745     \\
                             &                & Reply          & 86         & 163              & 95.9\%     & 4.1\%        & 342        & -0.918    & 0.868         & 4,316  & 12,397       & 0.668     \\
                             &                & Quote          & 109        & 199              & 92.8\%     & 7.2\%        & 787        & -0.856    & 0.818         & 6,315  & 26,025       & 0.682     \\ \cmidrule{3-13}
            Arson            &                & Hashtags (244) & 72         & 156              & 60.3\%     & 39.7\%       & 6,908,584  & -0.207    & 0.055         & 4,429  & 79,327,130   & 0.695     \\ \cmidrule{2-13}
            \ii{Category 1:} & ArsonEmergency & Retweet        & 493        & 592              & 99.5\%     & 0.5\%        & 6,645      & -0.989    & 0.988         & 12,076 & 21,526       & -0.787    \\
            \ii{Supporters}  &                & Mention        & 288        & 149              & 57.0\%     & 43.0\%       & 968        & -0.140    & 0.107         & 3,206  & 7,523        & 0.702     \\
            \ii{Category 2:} &                & Reply          & 247        & 105              & 70.8\%     & 29.2\%       & 476        & -0.417    & 0.196         & 2,041  & 3,031        & 0.606     \\
            \ii{Opposers}    &                & Quote          & 190        & 104              & 86.0\%     & 14.0\%       & 330        & -0.721    & 0.613         & 1,268  & 1,542        & 0.402     \\ \cmidrule{3-13}
                             &                & Hashtags (245) & 493        & 592              & 98.4\%     & 1.6\%        & 114,797    & -0.968    & 0.965         & 12,867 & 424,389      & 0.250     \\ \cmidrule{2-13}
                             & AFL            & Retweet        & 30         & 69               & 100.0\%    & 0.0\%        & 10         & -1.000    & 1.000         & 5,735  & 7,047        & 0.874     \\
                             &                & Mention        & 19         & 22               & 100.0\%    & 0.0\%        & 5          & -1.000    & 1.000         & 7,740  & 19,222       & 0.928     \\
                             &                & Reply          & 16         & 11               & 100.0\%    & 0.0\%        & 7          & -1.000    & 1.000         & 4,815  & 6,060        & 0.682     \\
                             &                & Quote          & 7          & 4                & --         & --           & --         & 0.000     & 0.000         & 1,821  & 1,670        & 1.000     \\ \cmidrule{3-13}
                             &                & Hashtags (338) & 42         & 73               & 100.0\%    & 0.0\%        & 97         & -1.000    & 1.000         & 11,573 & 81,457       & 0.949     \\
            \bottomrule
            
        \end{tabular}%
    } 
\end{table*}

Table~\ref{tab:summary_int_sig} shows that although the connectivity between the polarised groups was often statistically significant, it was often heterophilic rather than homophilic, meaning the groups often used the same hashtags. That said, there were large imbalances between the homophilic connections of the groups: YES accounts used YES-specific hashtags far more frequently than NO accounts used NO-specific hashtags in the Election dataset, while the same applied for Opposer accounts. In the other datasets, only Opposers’ use of Opposer-specific hashtags in the ArsonEmergency dataset stand out, and that is because there are so few connections, relatively (there were $7{,}875$ Opposer--Opposer connections, compared with $106{,}433$ Supporter--Supporter connections and $106{,}922$ Supporter--Opposer connections). Opposers strongly shared hashtags with Supporters, while Supporters also connected internally strongly to a similar degree. In all other cases, heterophilic connections dominated. This suggests that although the groups tended to interact amongst themselves, they often discussed similar topics, even with similar partisan leanings. A deeper exploration of which particular hashtags accounted for these heterophilic connections could reveal further insights regarding the axes of polarisation and agreement between the groups.

\begin{figure*}
	\centering
	\subfloat[SSM groups in Election]{
	    \includegraphics[height=0.22\textheight]{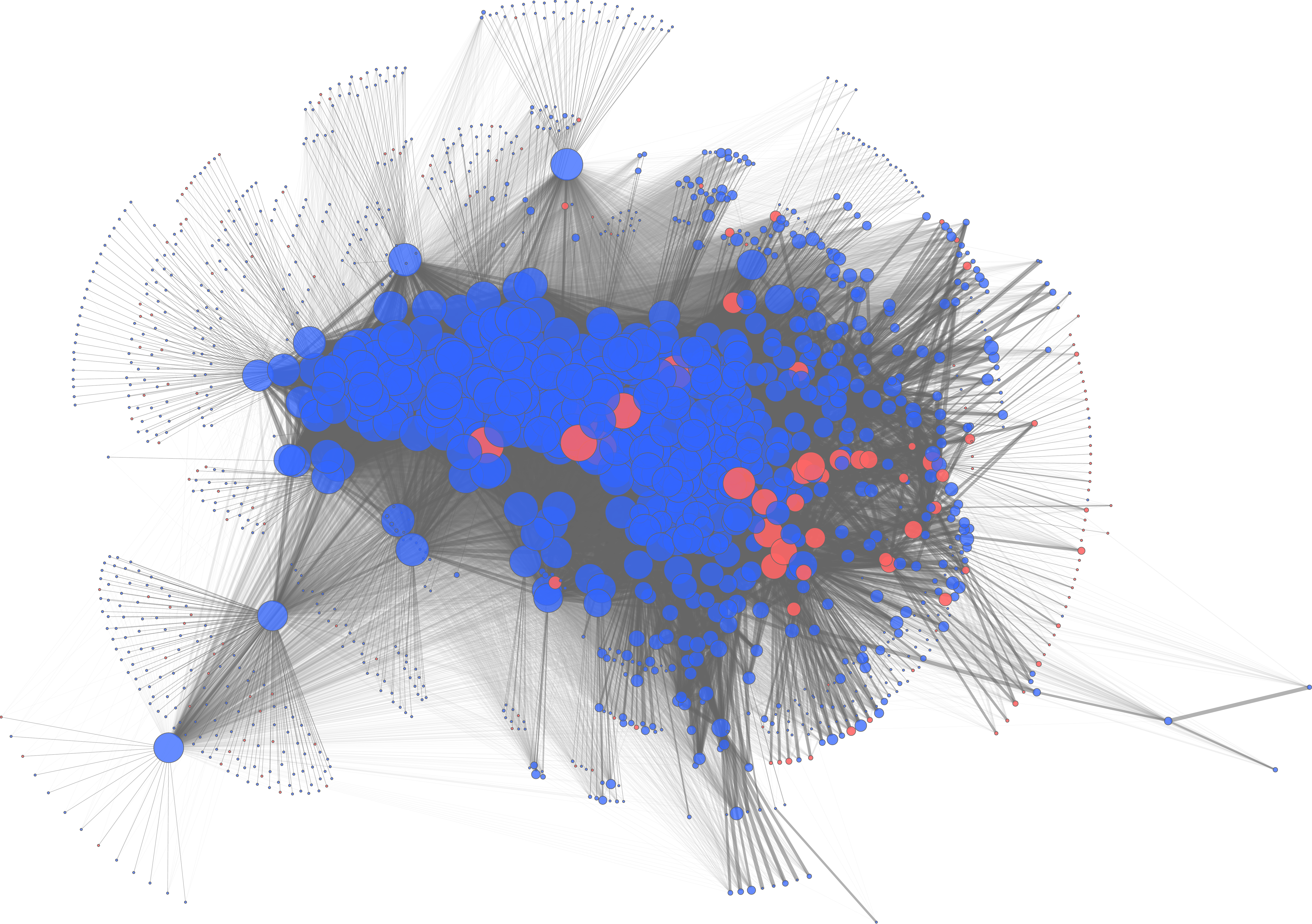}
	    \label{fig:ssm_groups_in_election_hashtag_co-mention_network} 
	} \hfill 
	\subfloat[Arson groups in Election]{
	    \includegraphics[height=0.22\textheight]{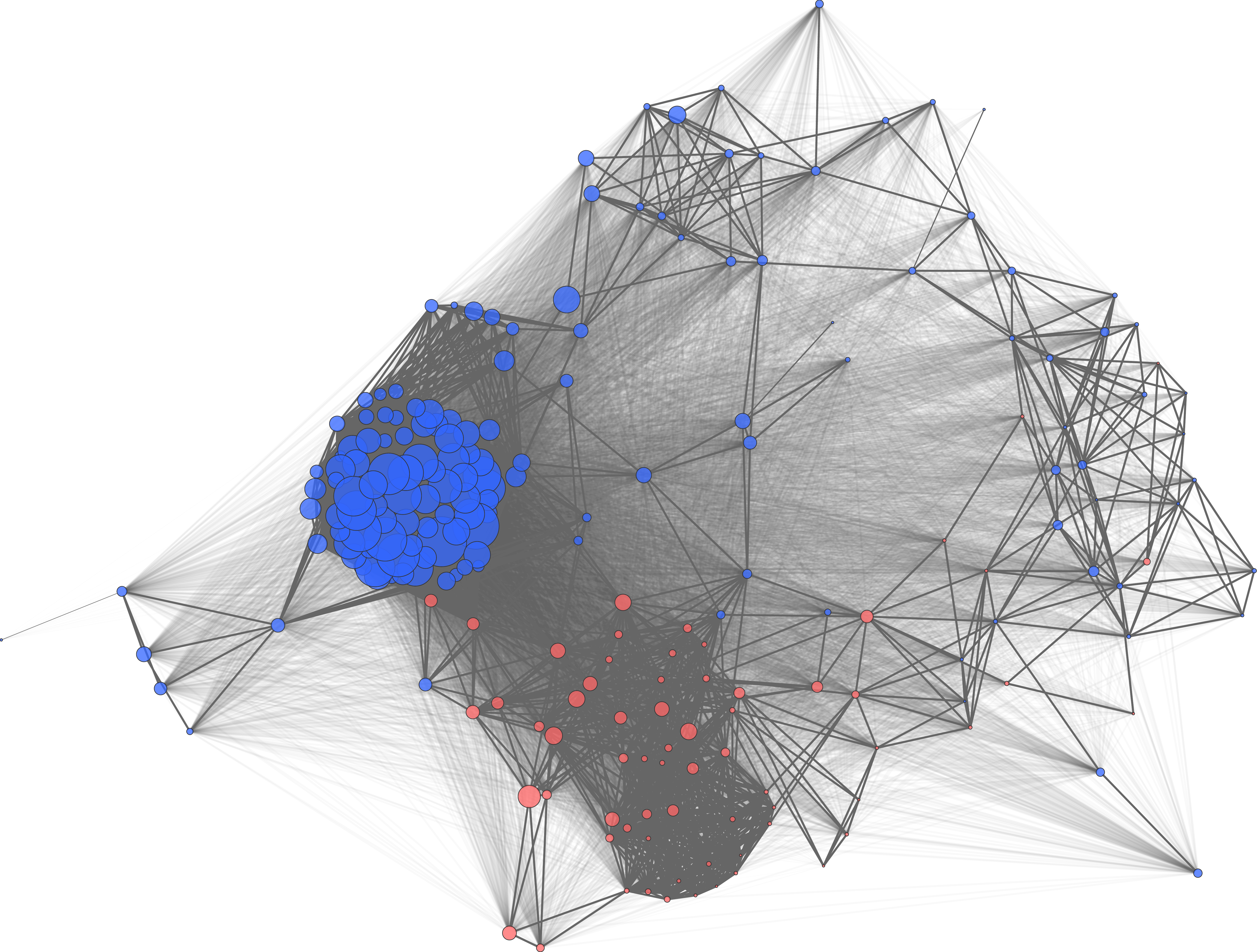}
	    \label{fig:arson_groups_in_election_hashtag_co-mention_network} 
	} \\
	\subfloat[SSM groups in ArsonEmergency]{
	    \includegraphics[height=0.22\textheight]{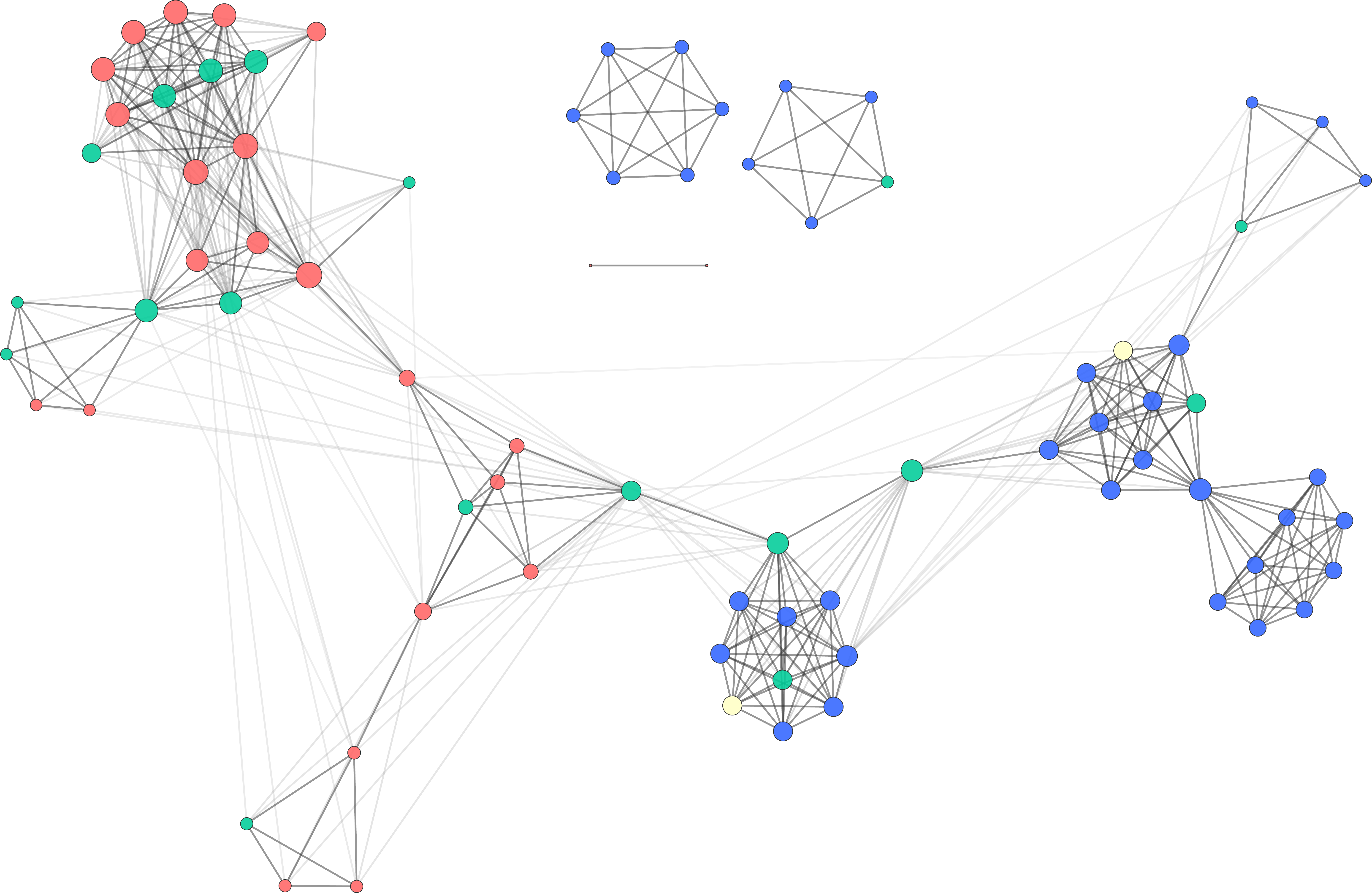}
	    \label{fig:ssm_groups_in_arson_hashtag_co-mention_network} 
	} \hfill 
	\subfloat[Arson groups in ArsonEmergency]{
	    \includegraphics[height=0.22\textheight]{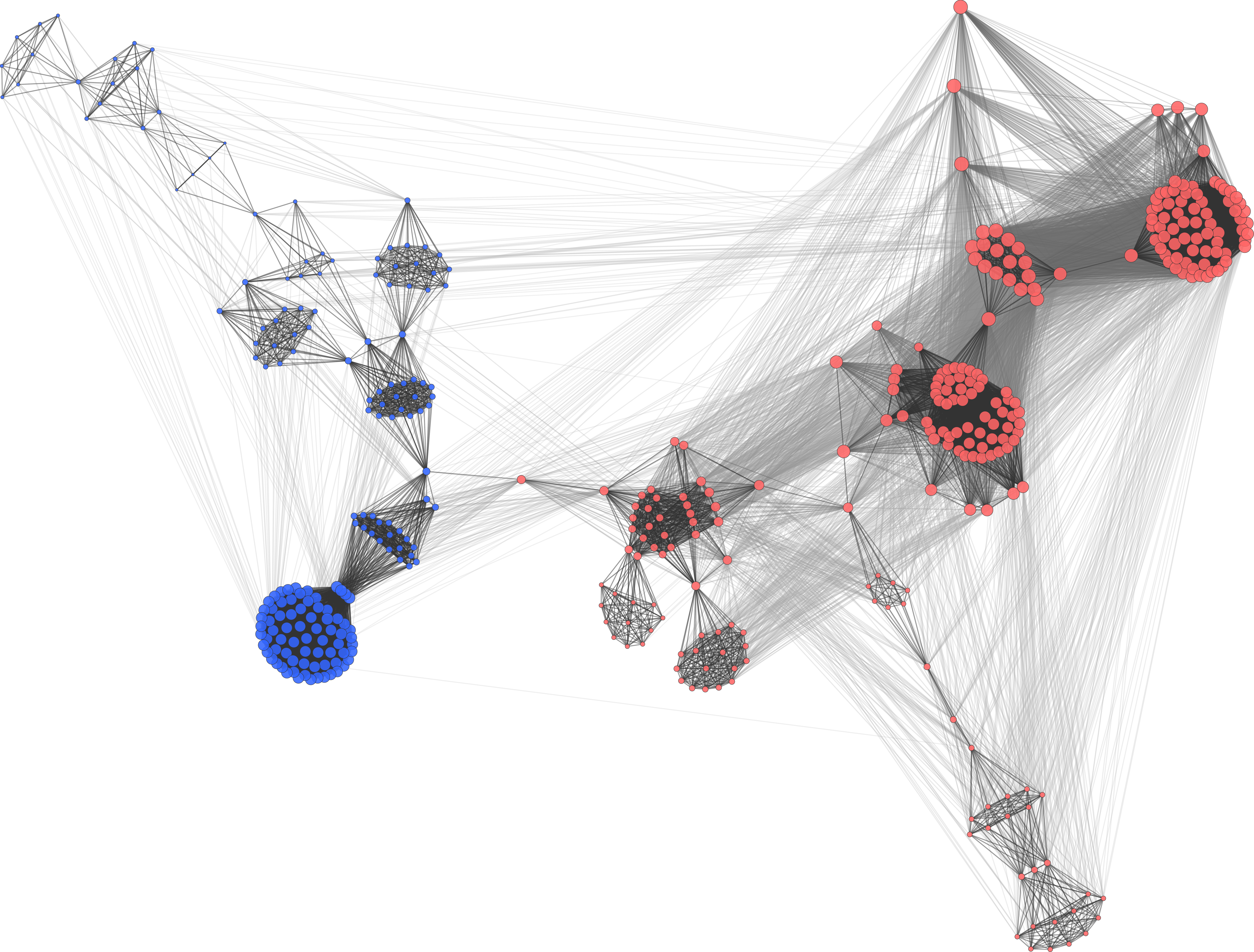}
	    \label{fig:arson_groups_in_arson_hashtag_co-mention_network} 
	} \\
	\subfloat[SSM groups in AFL]{
	    \includegraphics[height=0.22\textheight]{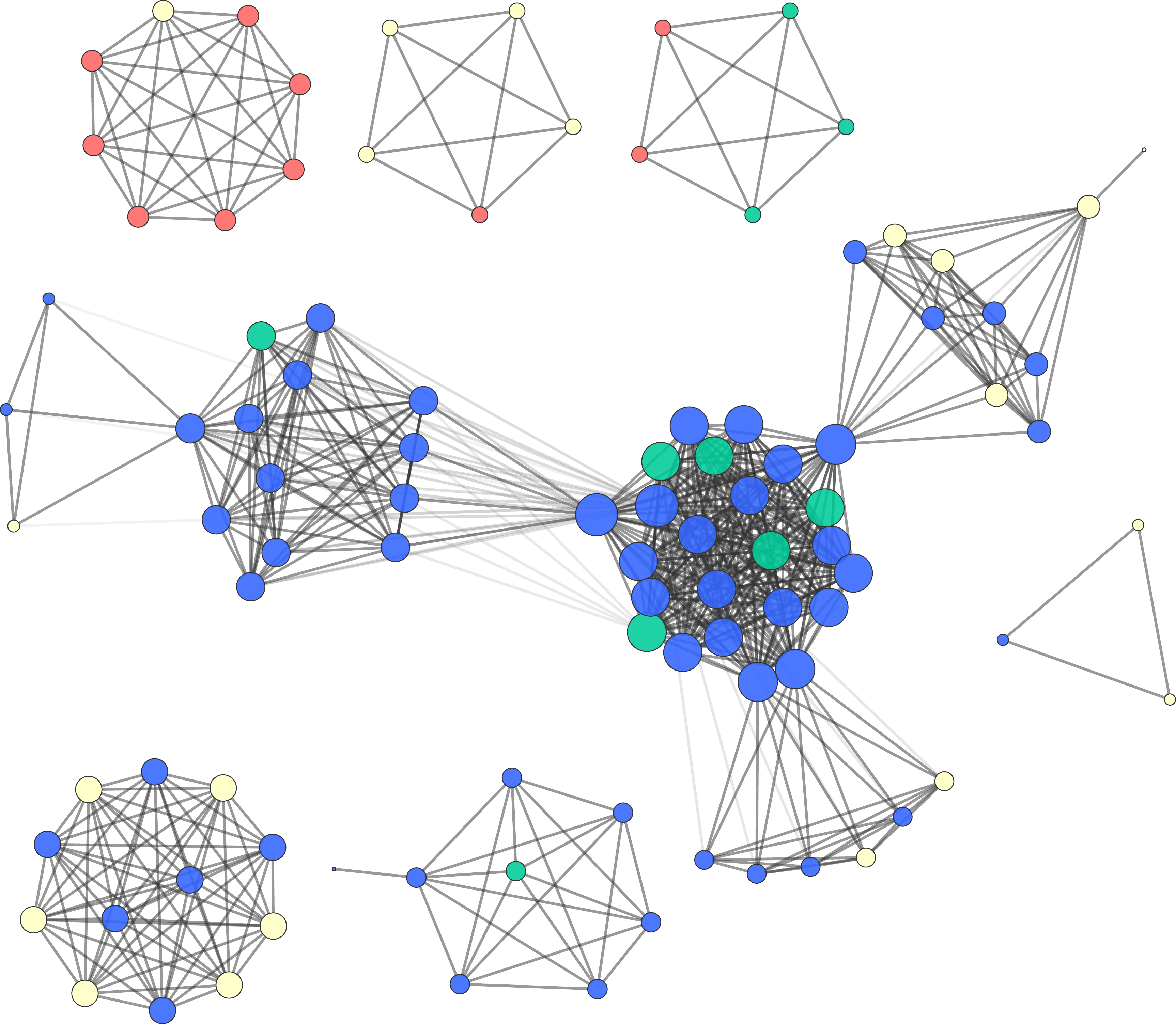}
	    \label{fig:ssm_groups_in_afl_hashtag_co-mention_network} 
	} \hfill 
	\subfloat[Arson groups in AFL]{
	    \includegraphics[height=0.22\textheight]{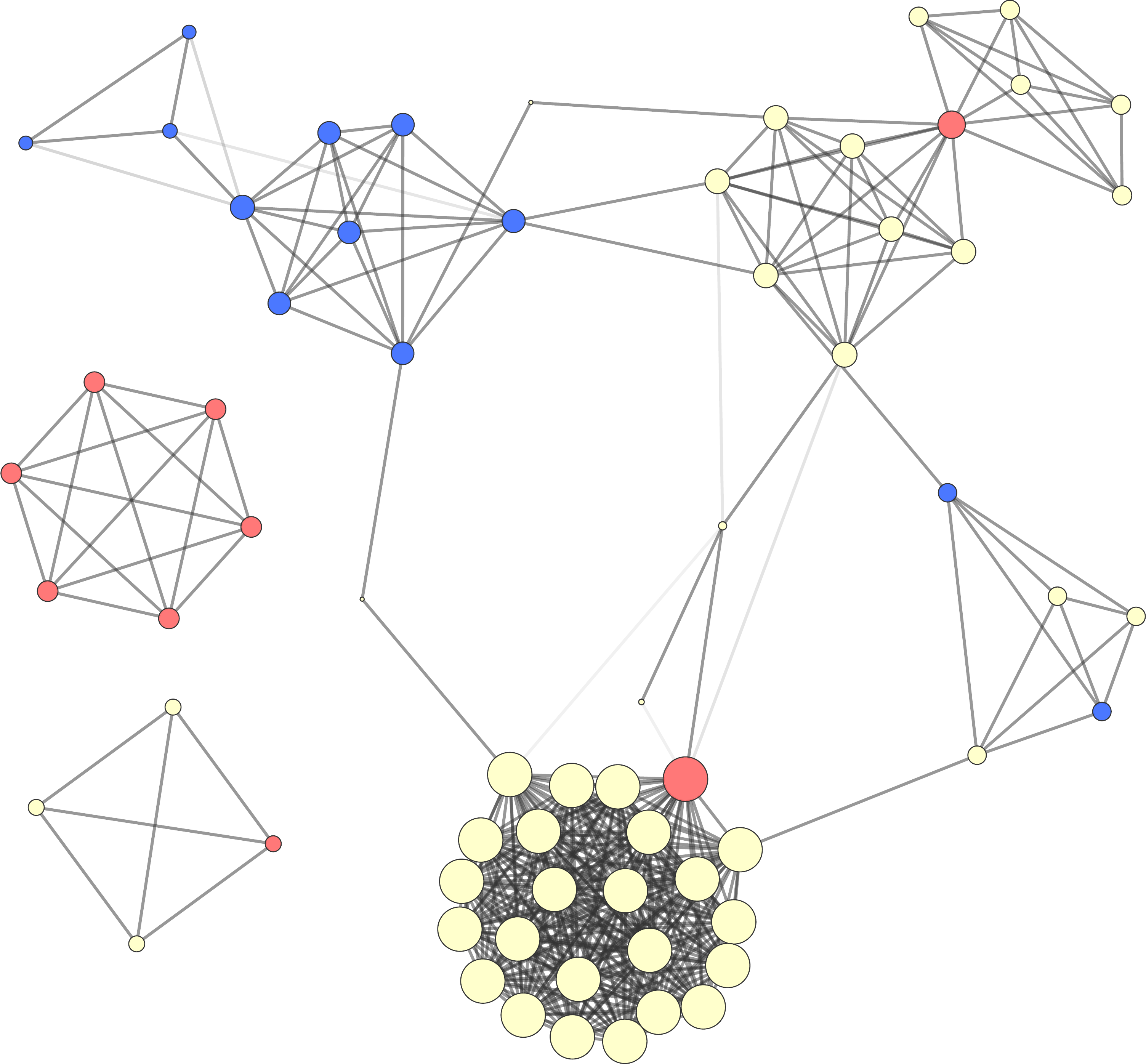}
	    \label{fig:arson_groups_in_afl_hashtag_co-mention_network} 
	}
	\caption{ \textbf{Polarisation in hashtag co-mention networks: }The largest components in hashtag co-mention networks of YES (blue) and NO (red) accounts and Supporter (red) and Opposer (blue) accounts active in the Election, ArsonEmergency and AFL dataset. Green nodes represent accounts that used both \hashtag{VoteYes} and \hashtag{VoteNo}, and yellow nodes represent OTHER nodes co-mentioning hashtags with affiliated accounts. Node size is determined by the sum of the backbone strength values on incident edges, i.e., degree weighted by backbone strength, indicating each node's embeddedness. Edge width describes the backbone strength. 
	}
	\label{fig:hashtag_co-mention_networks}
\end{figure*}

Using the backbone layout to visualise the hashtag co-mention networks (Figure~\ref{fig:hashtag_co-mention_networks}) makes clear the extent of the isolation of the groups despite their heterophilic connections, as well as the implications of the homophily measures. Nodes are sized according to 
weighted degree, using the backbone strength for edge weights. 
Edges are also coloured and sized according to backbone strength.

\subsection{Addressing the research questions}
We now directly consider the research questions posed in the Introduction.
\begin{description}
    \item[RQ1] \textit{Do Twitter accounts remain involved in Australian discussions for extended periods?} 
    
    We have shown that a significant number of accounts have remained active in the Australian Twittersphere over a number of years, with nearly 40\% of SSM YES accounts active nearly two years later in the lead up to the federal election and some hundreds still active in early 2020 during the Australian ``Black Summer'' bushfires. 
    Conversely, some hundreds of ArsonEmergency group members had also been active during the SSM discussion in $2017$, exhibiting a high degree of alignment with $65$ of $93$ Supporters using \hashtag{VoteNo} ($24$ of them also used \hashtag{VoteYes}) and $152$ of $240$ Opposers using \hashtag{VoteYes} ($33$ of them also used \hashtag{VoteNo}). 
    
    \item[RQ2] \textit{Is polarisation observed in one interaction type present across other interaction types?} 
    
    To the greater extent, the echo chambers observed in the Arson and SSM groups persisted through most interaction types according to E-I Index scores, at least moderately. E-I Index scores rose dramatically (towards heterophily) when the broader network was considered, indicating that the polarised groups were primarily polarised with regard to one another, and did, in fact, interact strongly with those outside their groups. 
    
    \item[RQ3] \textit{Do accounts found to be polarised in some discussions maintain their polarisation in different discussions, and does the theme of the discussion impact this polarisation?} 
    
    In contrast to the interaction networks, analysis of the common use of partisan hashtags revealed more heterophily in the Election dataset (in which a great variety of political issues were discussed), leading to the conclusion that although the groups mostly interacted amongst themselves, they discussed similar partisan topics and so probably also held similar positions on those topics (as described by the stance of the hashtags). The mix of political leanings exhibited by the NO accounts in the Election may have contributed to the overall greater heterophily in the Election dataset.
    
    In contrast to the Election, homophily remained very high in the topics discussed in the other datasets. For the ArsonEmergency discussion, this is likely due to the high alignment between the SSM and Arson groups, and for the AFL discussion, it is likely due to the match-specific nature of parts of the discussion.

    When considering the broader network, heterophily in discussions topics was mostly very strong across all datasets, except for when it was moderately homophilic and heterophilic amongst the SSM groups in the Election and the Arson groups in the ArsonEmergency datasets, respectively. This implies the SSM groups had their own distinct discussion topics during the election, which they shared amongst themselves but not with the broader community, and which might also provide an avenue for further integration.
    

\end{description}

\begin{figure}
    \centering
    \includegraphics[width=0.99\columnwidth]{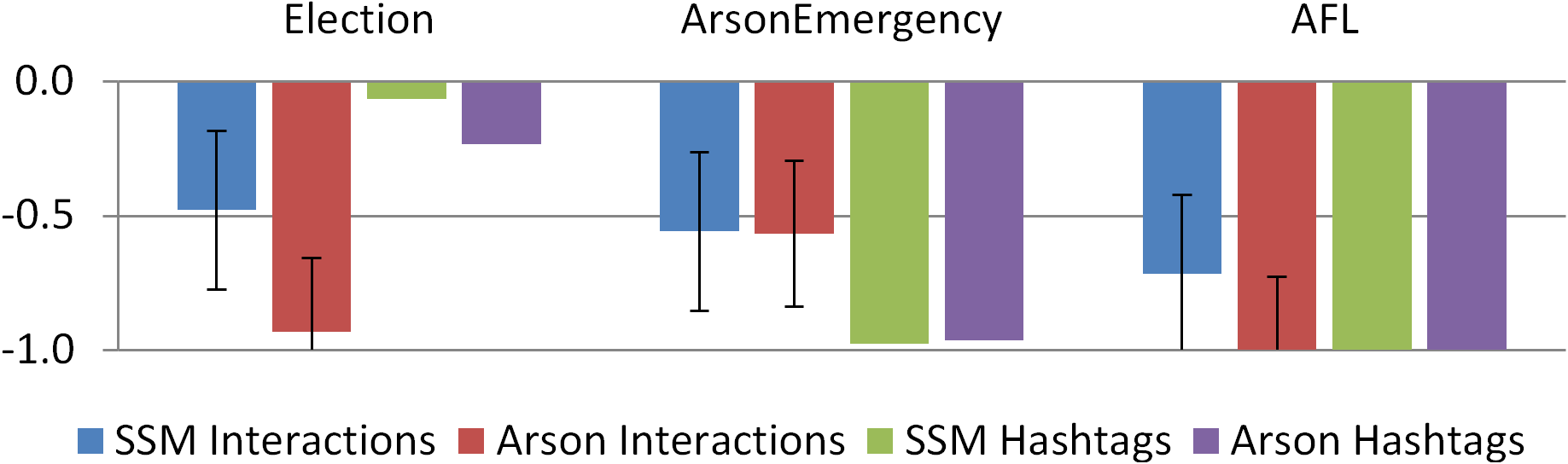}
    \caption{E-I Index scores for the SSM and Arson groups in each of the datasets. The interaction scores have been averaged, and the error bars indicate the standard deviation across all scores for that group. The hashtag scores refer to the hashtag co-mention networks. Scores approaching $-1.0$ indicate greater homophily, where $1.0$ indicates entirely heterophilic and $0$ indicates balance between homophilic and heterophilic edges (i.e., by the sum of their weights). As the hashtag bars represent single values, no error bars are required.}
    \label{fig:ei_index_summary}
\end{figure}

The interaction- and content-based E-I Index scores of the polarised groups revealed the groups interacted differently to how they discussed topics, raising the important question why.  
We summarise the E-I Index scores for each group and dataset in Figure~\ref{fig:ei_index_summary}, averaging the interaction network scores and contrasting them with the scores from the corresponding partisan hashtag co-mention networks. First, as mentioned above, polarisation varies from moderate to high across all interaction types for both SSM and Arson groups in all datasets, but is particularly pronounced between the Arson groups in the Election and AFL datasets. Second, the use of partisan hashtags during the Election was remarkably even compared to the other datasets. In fact, partisan hashtag use was almost entirely homophilic in the ArsonEmergency and AFL datasets. This could be explained in at least two ways. The first is that, although the partisan hashtags clearly align with political camps, the hashtags that co-occur with them in tweets might overlap significantly between the groups. Given the large number of them in the Election dataset ($200$), it is possible that there are many opportunities for accounts in different groups to use the same one. We might expect that, if this were the case, then the number of co-occurring hashtags in the other datasets should be low, however this is not what we find. Both had hundreds of co-occurring hashtags (see the counts next to the `Hashtags' labels in Table~\ref{tab:homophily_summary}). The second possibility is that the polarisation between the SSM and Arson groups is less to do with political opinions and more to do with social circles. People may have used \hashtag{VoteNo} in the SSM dataset for a variety of non-political reasons and factors, including religion, culture or general conservatism, and therefore may share the political opinions of many in the YES group. This orthogonality is perhaps less likely in the Arson group, given the motivation for being a Supporter or Opposer is easier to attribute to political outlook \cite{weber2020arsonemergency}, and we can see that the partisan hashtag use E-I Index score of the Arson groups in the Election dataset reflects this, being slightly more homophilic than that of the SSM groups.


\subsection{Addressing the hypotheses}
The statistical support shown in Table~\ref{tab:homophily_summary} for the hypotheses presented at the end of the Datasets section is mixed. 

SSM groups were very polarised in discussion topics in the ArsonEmergency and AFL datasets, but much less so in the Election dataset, and their interactions varied from moderately to highly homophilic (especially amongst the few present in the AFL dataset). In this way, the interactions observed indicate socially connected groups, while the content connections suggest they shared discussion topics strongly, even when they may have been partisan in nature.

Similarly, Arson groups interacted in strongly polarised ways in the Election and AFL datasets, but were only strongly homophilic in retweets and quotes in the ArsonEmergency dataset, where they were initially identified. Their connections were only weakly to moderately homophilic in their mentions and replies, respectively, in that dataset. Their use of content, however, was strongly polarised in all but the Election, where again they seemed to often share partisan discussion topics, but only within established social groups. The lower homophily in the reply and networks seems to suggest that Supporters and Opposers were willing to bridge the gap between the groups, but this may be a reflection of direct conflict, rather than genuine debate, based on the degree of aggressive behaviour observed in the tweets by \citeA{weber2020arsonemergency}.

\section{Discussion}

This work touches on a variety of research
questions, including how people decide their position in a social space when presented with
conflicting opinions about contentious topics, how political ideology drives people's stance
on issues, and what could make an echo chamber transient or persistent.
How behavior is affected by the social relations is described as one of the classic questions of social theory \cite{granovetter1985economic}. A listener who is not an active part of the conversation experiences the occurrences of the others' actions as ``events occurring in outer time and space'' \cite{garfinkel2005}. This view on shared events is a motivating factor for studying interactions which do not share physical presence, such as those in the online space. Studies have shown that people are influenced by online interactions, for instance, when it comes to making decisions about vaccination, opinions about vaccination on Twitter can act as a precursor to making a practical decision \cite{dunn2015associations}. 

Our analysis of the structural properties of a variety of networks based on their follower relations, interactions and hashtag use suggest that accounts expressing positive opinions about marriage equality (in the SSM dataset) or the arson narrative (in the ArsonEmergency dataset) were more closely connected in some parts of these networks leading to greater statistical homophily. Similar patterns held for those arguing against marriage equality and the arson narrative. 
A number of factors could be involved in causing this connection preference, some of which have been previously identified in the literature \cite{rogers1970homophily}. These include that communication is more effective amongst those who share common meanings, attitudes and beliefs.
Use of common information sources leads to a perception of greater trustworthiness and credibility within a community, while heterophilic interaction risks distortion of the message and potential for cognitive dissonance inasmuch as new messages can conflict with current beliefs. 
Such interactions can be valuable, however, helping to break the filter bubbles, exposing people to new ideas and points of view and challenging them to critically evaluate their own. 

Based on our observations, the primary cause for persistent polarisation may be the existence of social groups moreso than differences in opinions. As discussed above, a reason for homophilic connections is similarity between conversants, but that similarity may be due to being friends or acquaintances, rather than on less personal attributes, such as education or social status \cite<as noted elsewhere in the literature, e.g.,>{rogers1970homophily}. This is reinforced by the fact that the use of partisan hashtags in the Election dataset was so evenly distributed, suggesting that although the accounts interacted in what might be called echo chambers, they often discussed similar topics and held similar partisan views. In that sense, they may be more accurately described as social circles. That said, in other discussions, not only did they not interact, but they did not share content either, particularly in the ArsonEmergency dataset, so concerns that people are cutting themselves off from alternative viewpoints remain. Evidence from heterophilic connections in the ArsonEmergency dataset also aligns with observations of a high degree of antagonism \cite{weber2020arsonemergency}.

More broadly, Bruns' criticism of lack of clear definitions for the terms `echo chamber' and `filter bubble' \citeyear{Bruns2019} is well-founded, but these labels still hold value for communicating high level concepts. We offer a conceptual definition of an echo chamber as a community formed around a shared opinion on a particular issue or discussion topic, within which that same opinion is reinforced as part of the community's interactions and discussion. This is consistent with the literature \cite{garimella2018}. The members still interact with those outside the echo chamber, but may do so by also discussing other issues, 
which is in line with 
Georg Simmel's theory of intersecting `social circles'~\cite{simmel1908geheimnis}. Echo chambers can be identified as communities whose content, when analysed, is highly focused and of a similar opinion (e.g., through the use of partisan hashtags, which declare a stance on an issue), but whose members still interact frequently with those outside the community.
The members of a filter bubble, in contrast, lack significant interaction with those outside the community (i.e., instigated from within). This situation is often blamed on OSN recommendation algorithms in pursuit of personalised information offerings \cite{Pariser2012,Massanari2016,Bruns2019}. This kind of connectivity can be observed with network analysis and the discussion topics and stances can be identified with content analysis, but hard and fast rules such as `filter bubble members never interact with new content' are too strict to be of use in the highly varied world of social media. Of course, these definitions are limited to the OSNs (and other communication environments) available for analysis. A person might only ever tweet about arson, but will still interact with family, friends and workmates outside of Twitter, so a filter bubble is only likely to occur in the most extreme of circumstances (e.g., isolated cults).


\subsection{Critique and future work}
There are a number of ways to improve the approaches we have taken in this study, including the following considerations.
    
    The weights in the hashtag co-mention networks are calculated as the sum of the products of each pair of accounts' uses of a common hashtag, which may potentially inflate weights and not reflect imbalanced use between the members of the pair. Others \cite<e.g.,>{magelinski2021} have used the minimum instead, or a more sophisticated calculation may be warranted. 
    
    In fact, there may be benefit in additionally scaling edge weights by user activity: if an account is very active, they might co-use a hashtag more often just by chance, which will connect them to other users of the hashtag with very heavy edge weights. Taking relative activity into account may lighten these edges.
    
    The manner in which partisan hashtags are chosen is also, to some degree, a subjective activity. Furthermore, the faux partisan hashtags are highly likely to generate polarised groups, after all that is how they are chosen. We have, however, revealed interesting findings in the hashtags that co-occur with them, so there is merit in the approach but deeper investigation is required. For example, these commonalities could be studied separately by ignoring the specifically partisan hashtags as the hashtag co-mention network is created from the filtered tweets.

    The assumption that homophilic and heterophilic interactions are all equally representative of civil communication is lacking, and deeper examination is required to determine to what degree the interactions are positive or negative. \citeA{weber2020arsonemergency} found high degrees of aggression between the Arson groups, so it is possible that that aggression exists in the heterophilic connections in the other datasets. Methods exist for examining online group conflict that could be applied for this purpose \cite<e.g.,>{kumar2018conflict,DattaA19conflictnetwork}. An analysis of URL-sharing behaviour in these datasets may also reveal shared or divided stances on issues, as defined by what the URLs refer to.

\section{Conclusion}


Echo chambers on OSNs provide fertile ground for misinformation and polarisation on social and political issues, which can influence offline behaviour with real world effects such as vaccine hesitation and even violence. This study begins by identifying the SSM groups, a pair of polarised groups in the Twitter discussion surrounding the 2017 Australian postal survey on marriage equality. The activities of the SSM groups and of the previously identified Arson groups \cite{weber2020arsonemergency} are tracked over several Twitter datasets spanning a long period and a variety of discussion topics. The aim of the study has been to characterise the nature of their polarisation in terms of the interactions used and the topics discussed to determine if such communities are persistently polarised, or whether they mix over time as the issues at hand change.

Our findings reveal that persistent communities of Australian Twitter users exist and remain polarised in the social groups they form over periods of several years, but that the topics they discuss are often common, even in the context of partisan topics. Furthermore, these polarised groups interact strongly with those outside their groups even while they avoid each other, which offers hope that the echo chambers they form between themselves can be pierced and infiltrated through further encouraging and facilitating engagement with the broader online community.

%

\section{Copyright statement}
Please  be  aware that the use of  this \LaTeXe\ class file is
governed by the following conditions.

\subsection{Copyright}
Copyright \copyright\ \volumeyear\ SAGE Publications Ltd,
1 Oliver's Yard, 55 City Road, London, EC1Y~1SP, UK. All
rights reserved.

 \begin{acks}
This research was partially supported by Australian Research Council's (ARC) Centre of Excellence for Mathematical and Statistical Frontiers, Australia under project ID CE140100049.

 The authors also acknowledge support from the ARC's Discovery Projects funding scheme under project DP210103700.
 \end{acks}

\begin{ethics}
 All data was collected, stored and analysed in accordance with Protocols \#170316 and H-2018-045 as approved by the University of Adelaide's human research ethics committee.
\end{ethics}





\section{Supplementary material: Datasets}
\subsection{The marriage law postal survey (late 2017)}



In late $2017$, the Australian federal government conducted an optional national postal survey asking Australian voters ``Should the law be changed to allow same-sex couples to marry?''\footnote{\url{https://www.abs.gov.au/ausstats/abs@.nsf/mf/1800.0}} On the basis of a majority affirmative result, the government would commit to passing legislation to change the Marriage Act accordingly. From August, when the survey was announced, through to the final acceptance date of the ballots in November and beyond, discussions and debate raged on social media with strong opinions both for and against marriage equality. Ultimately, over $60\%$ of the nearly $13$ million responses voted `yes' and the Australian Parliament changed the law to permit marriage between any two individuals.

During three months of the campaign, we collected tweets from Twitter's $10\%$ academic sample stream\footnote{The Decahose: \url{https://developer.twitter.com/en/docs/twitter-api/enterprise/decahose-api/overview/decahose}} based on the keywords \hashtag{MarriageEquality},\footnote{All hashtag analysis was performed ignoring case, but capitals are included here for readability.} \hashtag{SSM}, \hashtag{auspol}, \hashtag{VoteYes}, and \hashtag{VoteNo}, capturing close to $80$k tweets (and associated metadata) by almost $55$k unique accounts. 

The hashtags used as keyword filters belonged to two categories: general marriage equality-related terms (\hashtag{MarriageEquality}, \hashtag{SSM}, and \hashtag{auspol}), and ones clearly reflecting an opinion (\hashtag{VoteYes} and \hashtag{VoteNo}).
We focused on the $17.3$k accounts which used the opinion-linked hashtags, which we hypothesised would have relatively high structural cohesion around users of the same hashtag, and low structural cohesion among users of different hashtags. 
YES accounts were those that used only \hashtag{VoteYes}, NO accounts used only \hashtag{VoteNo}, and BOTH accounts used both hashtags. 
Of these, there were slightly more YES accounts than NO accounts ($8.6$k to $7.9$k), and those using both made up just under $5\%$ of the accounts using opinion hashtags ($778$). YES accounts contributed more tweets ($18{,}621$) than NO accounts ($11{,}261$) and BOTH accounts ($7{,}246$).

\begin{figure*}[ht]
	\centering
    \includegraphics[width=0.99\textwidth]{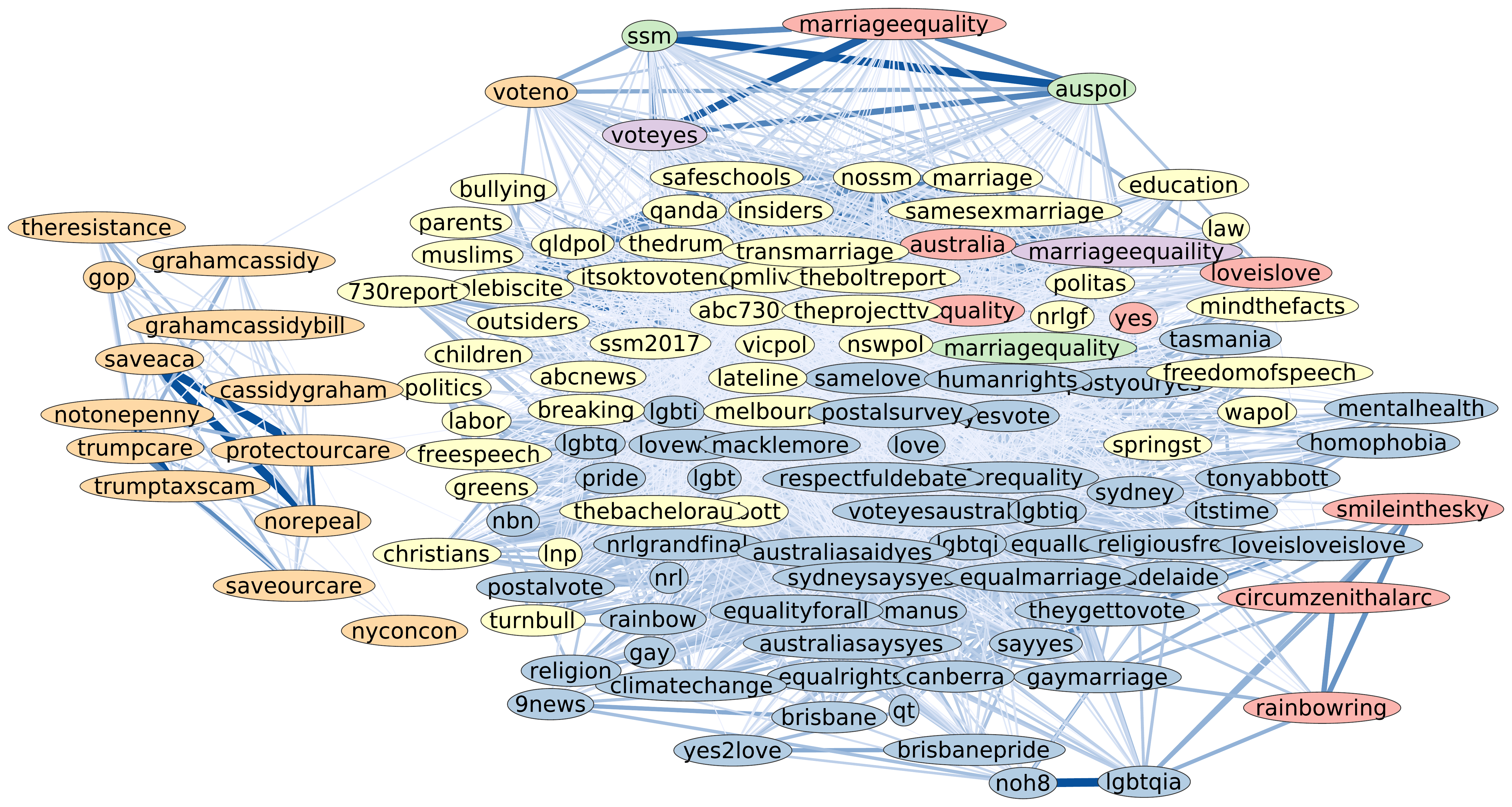} 
    \caption{The hashtag network from the SSM dataset. Two hashtag nodes are linked if they were tweeted by the same user (though not necessarily in the same tweet), and the size and colour of the edge represents the frequency of co-mentioning (wider and darker = more frequent). Nodes are coloured according to Louvain cluster. Names of prominent public figures have not been anonymised in order to provide context. The orange cluster on the left clearly refers to US politics rather than the Australian SSM postal survey. \label{fig:simplifiedhashtags}}
\end{figure*}

Some cleaning of the data was required due to international overlap with \hashtag{VoteNo}, which was also used in American discussions surrounding a medical insurance-related bill before the US Congress at the time. 
These tweets were identified through the use of a hashtag network. 
The network is visualised in Figure~\ref{fig:simplifiedhashtags} with a force-directed layout clearly showing a minimally linked cluster of hashtags on the left that relate to the foreign discussion. $6{,}295$ tweets posted by $5{,}366$ accounts mentioning the hashtags in the orange coloured Louvain cluster \cite{blondel2008} to the left (other than \hashtag{VoteNo}) were identified as pollution and removed.

This dataset is referred to herein as the \emph{SSM} dataset, and the YES and NO accounts as the \emph{SSM} accounts (BOTH accounts are not included in the analysis as their position on the matter is just as obscure as OTHER accounts).

\subsection{The Australian federal election (May 2019)}\label{sec:electionsData}


A total of $4{,}429$ of the YES, NO and BOTH accounts ($3{,}390$, $631$ and $408$, respectively) were active during the election period surrounding the Australian federal election held on the 18th of May, 2019. Their activity was tracked, 
resulting in a dataset of nearly $400$k tweets spanning three weeks. These activities were obtained, post-election, by retrieving their timelines via Twint\footnote{\url{https://github.com/twintproject/twint}} (a tool that obtains Twitter data directly from its web UI avoiding any recommender influence or constraint present in the APIs). Nearly $3.4$k YES accounts were active during the campaign, compared to only $631$ NO accounts. The data includes a variety of politically-relevant hashtags, and in particular we have identified $44$ partisan hashtags. 
   

\subsection{Australia's ``Black Summer'' (2019-2020)}\label{sec:bushfiresData}
During the 2019-2020 southern summer, referred to as Australia's `Black Summer', bushfires burnt over $16$ million hectares of the Australian mainland, destroyed over $3{,}500$ homes, and caused at least $33$ human and a billion animal fatalities \cite{nsw_bushfire_inquiry2020}. 
While scientists attributed these bushfires to natural causes such as lightning, an alternative theory labelled arson as the cause of bushfires. At the peak of the bushfires season the hashtag \hashtag{ArsonEmergency} started trending on Twitter and was observed to include a high proportion of bot and troll activity \cite{Stilgherrian2020zdnet,GrahamKeller2020conv}. 
\citeA{weber2020arsonemergency} collected a dataset of tweets during that period, both before and after news of bots and trolls reached the mainstream media. 
The dataset consisted of $27.5$k 
tweets containing the term `ArsonEmergency' posted by $12.9$k 
unique accounts over $18$ days in early January 2020. 
The Tweets were obtained with Twitter's Standard Search API using Twarc.\footnote{\url{https://github.com/DocNow/twarc}} \citeA{weber2020arsonemergency} found two polarised communities in the retweet network, which we refer to here as the Arson groups. One community strongly supported the arson narrative (\emph{Supporters}), claiming arson was the cause of the bushfires, posting $6{,}972$ tweets, while the other community opposed that narrative with fact-check articles and official announcements in $3{,}587$ tweets (\emph{Opposers}). A second study on this hashtag and contemporary news media reports found evidence of a disinformation campaign conducted by trolls, which appeared coordinated with the help of prominent public figures \cite{keller2020arson}.

This dataset provides our second set of polarised accounts.


\subsection{AFL (March 2019)}\label{sec:aflData}


A further, non-political dataset that could also exhibit patterns of polarisation was sought as a contrast. 
Australian Rules Football is a national pastime in Australia, particularly following the national competition run by and synonymous with the Australian Football League (AFL). A three-day collection of AFL discussions was conducted over a weekend in March, 2019, just as the annual season began \cite<previously published in>{WeberNMF2021reliability}. Although a federal election was expected around this time, it was not called for another two weeks, and so little political content was expected to be captured. The collection tool RAPID \cite{rapid2017} was used to stream all tweets from the Standard Twitter v1.0 Streaming API (up to rate limits) using the keyword `afl' and a language filter for English and undefined (i.e., a `lang' value of `und', which captures text too short to inform Twitter's language detection).

\subsection{Polarisation labelling}

\begin{figure}[th]
    \centering
    \subfloat[The SSM groups.]{
        \includegraphics[height=0.18\textheight]{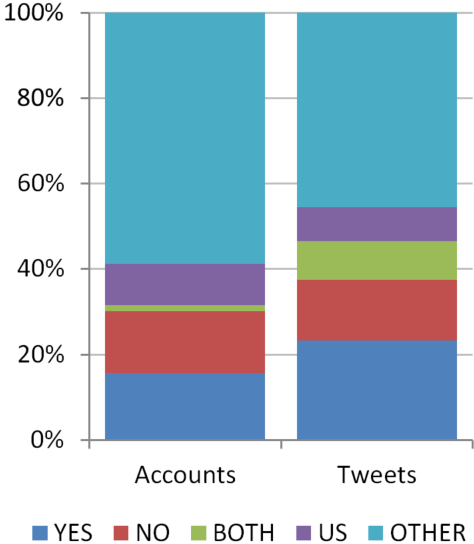}
        \label{fig:ssm_relative_contributions}
    } \hfill
    \subfloat[The Arson groups.]{
        \includegraphics[height=0.18\textheight]{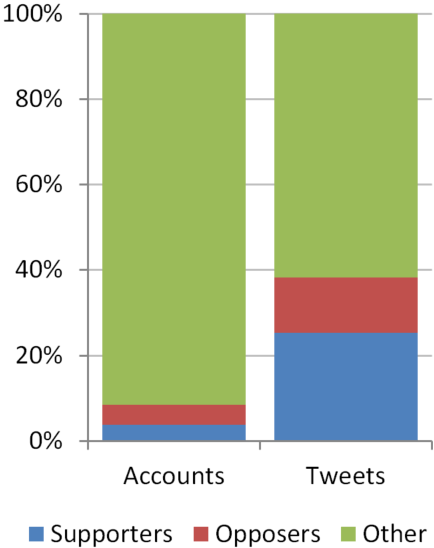}
        \label{fig:arson_relative_contributions}
    }
    \caption{The relative contributions of the polarised SSM and Arson groups in the datasets in which they were found.}
    \label{fig:ssm_arson_relative_contributions}
\end{figure}

Different methods were used to identify polarised communities in the SSM and Bushfires datasets due to the different ways in which they were collected. The sizes of the groups discovered are shown in Table~\ref{tab:polarised_group_stats} and their relative contributions in Figure~\ref{fig:ssm_arson_relative_contributions}. 


Generalising our terminology, we refer to YES and Supporter groups as Category 1 accounts and NO and Opposer groups as Category 2 accounts later in this work. Any unaffiliated accounts appearing in networks are given the label OTHER.

\begin{table}[ht]
    \centering
    \caption{Sizes of the labelled polarised communities.}
    \label{tab:polarised_group_stats}

        \begin{tabular}{@{}rrrrr@{}}
            \toprule
            \multicolumn{3}{c}{SSM} & \multicolumn{2}{c}{ArsonEmergency} \\ 
            \cmidrule(r){1-3}      \cmidrule(l){4-5}
            YES    & NO     & BOTH  & Supporters        & Opposers       \\
            8,623  & 7,880  & 778   & 497               & 593            \\ 
            \bottomrule
        \end{tabular}%



\end{table}

A summary of the content of the datasets and the extent of the polarised group presence in them in shown in Table~\ref{tab:dataset_stats}. 

\begin{table*}[ht!]
    \centering
    \caption{Relevant statistics of the datasets analysed.}
    \label{tab:dataset_stats}
        \begin{tabular}{@{}l|rrrr|rrrrr@{}}
            \toprule
            Dataset        & Retweets & Mentions & Replies & Quotes & Accounts & YES   & NO  & Supporters & Opposers \\
            \midrule
            Election       & 331,682  & 51,673   & 12,397  & 26,025 & 4,429    & 3,390 & 631 & 72         & 156      \\
            ArsonEmergency & 21,526   & 7,523    & 3,031   & 1,542  & 12,872   & 698   & 148 & 493        & 592      \\
            AFL            & 7,047    & 19,222   & 6,060   & 1,670  & 11,573   & 376   & 53  & 42         & 73       \\
            \bottomrule
        \end{tabular}%
\end{table*}

\section{Supplementary material: Results}

\subsection{Networks based on interactions}
The results of a systematic examination of the presence and interactivity between the YES and NO and Supporter and Opposer accounts in the AFL, Election and ArsonEmergency datasets are presented in Table~\ref{tab:summary_int_sig}. For each group and interaction in each dataset, we considered how often they interacted amongst themselves (i.e., homophilic connections) and with each other (i.e., heterophilic connections, ignoring the broader network). For both circumstances, we present the sum of the edge weights rather than just the edge count. We used binomial tests to examine the null hypothesis that they had no connection preference, and where $p$-values are presented, this is the confidence in rejecting the null hypothesis. This provides us with evidence to address all three research questions. 

\subsubsection{Polarisation on \hashtag{ArsonEmergency}.} 
Given polarisation between Supporters and Opposers was first observed in the ArsonEmergency dataset, we can now see that that polarisation extends across the other interactions significantly in all cases except for Opposers’ use of mentions. Weber's findings that Opposers focused almost exclusively on retweets, while Supporters also made use of many other interactions is corroborated – in fact, it is clear that Supporters’ use of non-retweet interactions was less polarised that it might have been, as they interacted with Opposers between 20\% and 35\% of the time.

SSM accounts were fewer in number and less active in the ArsonEmergency dataset, and only statistically significantly maintained their polarisation in the retweet network and only amongst NO accounts in the quotes network. Considering the raw numbers, we see NO accounts also used mentions and replies more often than YES accounts, but were relatively balanced in how they used them. In contrast YES accounts connected almost exclusively to other YES accounts with the same interactions.

\subsubsection{Polarisation leading up to the federal election.}
The polarisation in the Election dataset is statistically significant across all groups and interactions, and it is homophilic in all but one condition: NO accounts mention YES accounts more frequently than other NO accounts. This is not necessarily surprising given there are more than five times more YES accounts than NO accounts active in the dataset. 

Amongst the Arson groups, in particular, the echo chamber effect (with relation to each other, at least) is stark, with both groups preferring internal to external connections by several orders of magnitude. The smallest ratio of homophilic to heterophilic connections was Supporters' use of quotes (at $\approx 6.8$), but most were much greater than that. This marked polarisation is also immediately apparent in visualisations of the interaction networks (Figures~\ref{fig:arson_groups_in_auselecvote_retweet_network} to~\ref{fig:arson_groups_in_auselecvote_quote_network}).

The results amongst SSM groups were also all statistically significant, and in all but one case were homophilic, as mentioned above, but the pattern of polarisation differs due to the relative sizes of the groups present (Figures~\ref{fig:ssm_groups_in_auselecvote_retweet_network} to~\ref{fig:ssm_groups_in_auselecvote_quote_network}). The imbalance in use of interactions is immediately apparent, with the greater number of active YES accounts (presented in Table~\ref{tab:everpresent_accounts}) contributing more proportionally than NO accounts across all interaction types. YES accounts outnumber NO accounts five to one, but posted 8.2 times as many retweets, 10.5 times as many mentions, 8.4 times as many replies and 6.9 times as many quotes, so more YES accounts were present in the Election dataset but they were also more active. Furthermore, their echo chamber effect was more pronounced, retweeting, mentioning, replying to and quoting each other over 95\% 
of the time, while NO accounts interacted with each other slightly less than half the time. 
These findings indicate that: 1) polarisation detected amongst one type of interaction can be present across other types of interaction; and 2) polarisation detected in one issue-related discussion can be found in other issue-related discussions, including across a variety of interactions. 

The issues discussed in the ArsonEmergency and Election datasets can be regarded as at least partially political in nature, so the question remains whether the above phenomena persist in non-political discussions. We use the AFL dataset for this contrast, assuming that, whatever their political opinions, any alignment with people's political opinions is likely to be coincidental. Political discussion in the AFL dataset is minimal, and even the most prominent political hashtag is the non-partisan \hashtag{auspol}.

\subsubsection{Polarisation discussing the AFL.}
Very few Arson accounts interacted with other accounts in the AFL dataset, but where they did it was strongly homophilic relative to each group. The majority of their connections were to the broader network as sources of interactions (i.e., they reached out to others). 

SSM accounts also interacted rarely in the AFL dataset, but the much greater number of YES accounts were strongly homophilic in the connections they made, with respect to the two groups. Again, both groups interacted strongly with the broader network, with some accounts frequently the recipient of interactions rather than just the instigator, as was the case with the Arson group members.

\subsubsection{Summary of interaction network findings.}
In almost all circumstances, the echo chamber effect appears to be maintained to some degree, with internal connections preferred over external ones, especially between Supporters and Opposers. The only circumstance where that effect is reduced is in the NO groups' use of replies and mentions and Opposers' use of mentions in the ArsonEmergency dataset, where they more even in their connections. It is possible that some of these mentions were used for aggressive, rather than collegiate, interactions, but analysis of their content is required for this judgement and there were relatively few of these interactions, so any such judgement is unlikely to be indicative of a broader pattern of behaviour.

\subsection{Networks based on content}
Results so far indicate the echo chamber effect is strongly maintained across most interactions in most datasets, especially where there is reasonable amount of activity. Here we consider whether the topics also under discussion also exhibit similar patterns of polarisation, and we use hashtags as an indicator of those topics.

First, however, we must cull the hashtags under consideration, as the high frequency of popular hashtags can hamper the discovery of the structures underlying their use. Instead, as discussed above, we explicitly filter the most frequent hashtags and we additionally make use of partisan and faux partisan hashtags. Examining the distributions of hashtag use in each of the dataset revealed that removing the ten most frequent hashtags in each would be sufficient to avoid the majority of their binding effects (shown as the dashed red vertical lines in Figure~\ref{fig:hashtag_distributions_and_cutoffs}).


Second, we developed the (faux) partisan hashtag sets. In the Election dataset, we identified $44$ hashtags of the $200$ most frequent as clearly partisan (e.g., \hashtag{corrupt<party>} or \hashtag{<party>liars}). For the AFL and ArsonEmergency datasets, we identified the ten most frequently used hashtags unique to each group. We considered the tweets containing these hashtags and created hashtag co-mention networks using all the hashtags that appeared in them (save for the most frequently occurring hashtags, as mentioned above). The number of hashtags considered for each group and dataset is shown in parentheses next to the ``Hashtags'' label in Table~\ref{tab:homophily_summary}, which also shows the number of SSM and Arson group accounts present in the resulting networks, and their respective connectivity.

Above, Table~\ref{tab:summary_int_sig} shows that although the connectivity between the polarised groups was often statistically significant, it was often heterophilic rather than homophilic, meaning the groups often used the same hashtags. That said, there were large imbalances between the homophilic connections of the groups: YES accounts used YES-specific hashtags far more frequently than NO accounts used NO-specific hashtags in the Election dataset, while the same applied for Opposer accounts. In the other datasets, only Opposers’ use of Opposer-specific hashtags in the ArsonEmergency dataset stand out, and that is because there are so few connections, relatively (there were $7{,}875$ Opposer--Opposer connections, compared with $106{,}433$ Supporter--Supporter connections and $106{,}922$ Supporter--Opposer connections). Opposers strongly shared hashtags with Supporters, while Supporters also connected internally strongly to a similar degree. In all other cases, heterophilic connections dominated. This suggests that although the groups tended to interact amongst themselves, they often discussed similar topics, even with similar partisan leanings. A deeper exploration of which particular hashtags accounted for these heterophilic connections could reveal further insights regarding the axes of polarisation and agreement between the groups.

\subsubsection{Visualisation reveals deeper community structures.}
Using the backbone layout to visualise the hashtag co-mention networks (Figure~\ref{fig:hashtag_co-mention_networks}) makes clear the extent of the isolation of the groups despite their heterophilic connections, as well as the implications of the homophily measures. Nodes are sized according to 
weighted degree, using the backbone strength for edge weights. 
Edges are also coloured and sized according to backbone strength.

The relatively low homophily of the YES and NO groups during the Election (Figure~\ref{fig:ssm_groups_in_election_hashtag_co-mention_network}) is primarily due to the relatively small number of NO-only connections (see Table~\ref{tab:summary_int_sig}), which is evident from the NO nodes' dispersed placement throughout the network. Despite the placement, their size indicates they have high centrality and are therefore deeply embedded in the network. In contrast, the Supporter nodes active in the Election in Figure~\ref{fig:arson_groups_in_election_hashtag_co-mention_network} are not deeply embedded in the network (according to their sizes) but they clearly form a cluster of their own (to the bottom of the figure). The majority of Opposer nodes reside in a large cluster (top left) and \emph{are} deeply embedded. The relatively moderate E-I Index and assortativity scores in Table~\ref{tab:homophily_summary} ($-0.207$ and $0.055$, respectively) indicate that the Supporter nodes are highly connected to the Opposer nodes, which outnumber them, one to two ($72$ to $156$).

Both SSM and Arson groups formed mostly homophilic tight clusters in the ArsonEmergency dataset (Figures~\ref{fig:ssm_groups_in_arson_hashtag_co-mention_network} and~\ref{fig:arson_groups_in_arson_hashtag_co-mention_network}, respectively), but NO accounts were more often associated with BOTH accounts, which suggests they shared views on the arson narrative, given the alignment between NO and Supporter groups mentioned previously. Opposers and Supporters formed multiple separate clusters, but the most deeply embedded Opposers are clearly strongly concentrated in a single cluster (bottom left), while the deeply embedded Supporters form several groups. Deeper analysis is needed to examine which hashtags bound each different cluster.

Similar patterns of hashtag co-use are present in the AFL dataset (Figures~\ref{fig:ssm_groups_in_afl_hashtag_co-mention_network} and~\ref{fig:arson_groups_in_afl_hashtag_co-mention_network}), but unaffiliated accounts contributed more structure. The nature of the AFL discussion is relatively clustered in general, however, as sports fans discuss specific games, each of which has their own hashtag, which they use along with the \hashtag{afl} hashtag -- the top five used hashtags after \hashtag{afl} were \hashtag{aflPiesCats}, \hashtag{aflDogsSwans}, \hashtag{aflLionsEagles}, \hashtag{aflFreoNorth} and \hashtag{aflDeesPower}, all of which refer to the AFL and two teams that played each other in that round of the competition. It is therefore unsuprising to see some significant degree of clustering in these networks, but the fact that accounts from different groups do not seem to mix in each is notable, and may suggest a strong degree of influence from social circles.

\subsection{Homophily measures and the broader network}
The homophily measures in Table~\ref{tab:homophily_summary} provide a more nuanced view of the groups' homophily or heterophily in different circumstances than the statistics in Table~\ref{tab:summary_int_sig}. The SSM groups remained moderately to strongly polarised among all interactions except for mentions in the Election and ArsonEmergency datasets and the few quotes they posted in the AFL dataset. The Arson groups were mostly highly polarised in all cases except in replies (moderately) and mentions (mildly) in the ArsonEmergency dataset. Regarding their content, polarisation remained in the ArsonEmergency and AFL datasets but was only mild to moderate (E-I Indexes of $-0.066$ to $-0.207$ for SSM and Arson groups, respectively) during the Election.

Considering the broader network (as defined above), it is clear that all groups interacted and shared discussions with adjacent non-polarised accounts in all but one circumstance. Interestingly, the homophily in topics discussed increased modestly for the SSM groups in the Election, suggesting some further divide in the extra $4$k OTHER accounts.


\bibliographystyle{apacite}
\bibliography{references-new}

\end{document}